\documentclass[11pt, a4paper]{article}

\pdfoutput=1

\usepackage{amsmath}
\usepackage{amsfonts}
\usepackage{amssymb}
\usepackage{bbm}
\usepackage{verbatim}
\usepackage{dsfont}
\usepackage{booktabs}
\usepackage{slashed}
\usepackage{nicefrac}

\usepackage{epsfig}
\usepackage{color}
\usepackage[table]{xcolor}
\usepackage{graphicx}
\usepackage[font=small]{caption}
\usepackage[listofformat=empty,subrefformat=empty]{subfig}

\usepackage{float}
\usepackage{overpic}
\usepackage{tikz}
\usepackage{tikz-3dplot}
\usetikzlibrary{calc}
\usetikzlibrary{decorations} %
\usepackage{tikz-cd} 

\usepackage{bbold}

\usepackage{enumerate}
\usepackage{hhline}
\usepackage{multirow}

\usepackage[nosort]{cite}
\usepackage{xspace}
\usepackage{setspace}

\numberwithin{equation}{section}

\usepackage[pdftitle={Higher-Form Symmetries and their Anomalies in M-/F-theory Duality},
  pdfauthor={Mirjam Cvetic, Markus Dierigl, Ling Lin, Hao Y. Zhang},
  pdfsubject={M-theory, F-theory, generalized symmetries, anomalies},
  bookmarksopen, bookmarksnumbered, bookmarksopenlevel=2, colorlinks=false, linkcolor=blue, citecolor=blue, urlcolor=blue]{hyperref}

\renewcommand{\tfrac}{\genfrac{}{}{}1}
\newcommand{\bbZ}{\ensuremath{\mathbb{Z}}}
\newcommand{\bbC}{\ensuremath{\mathbb{C}}}
\newcommand{\bbP}{\ensuremath{\mathbb{P}}}
\newcommand{\bbF}{\ensuremath{\mathbb{F}}}

\begin{document}

\baselineskip=18pt

\vspace*{-2cm}
\begin{flushright}
	\texttt{CERN-TH-2021-091} \\
	\texttt{UPR-1311-T}\\
  \texttt{LMU-ASC 17/21}
\end{flushright}

\vspace*{0.6cm} 
\begin{center}
{\LARGE{\textbf{Higher-Form Symmetries and Their Anomalies\\in M-/F-Theory Duality}}} \\
 \vspace*{1.5cm}
Mirjam Cveti{\v c}$^{1,2,3}$, Markus Dierigl$^4$, Ling Lin$^{5}$, Hao Y.~Zhang$^{1}$\\

{
 \vspace*{1.0cm} 
{\it ${}^1$ Department of Physics and Astronomy, \\University of Pennsylvania,
Philadelphia, PA 19104, USA\\ \vspace{.2cm}
${}^2$ Department of Mathematics, \\University of Pennsylvania,
Philadelphia, PA 19104, USA\\ \vspace{.2cm}
${}^3$ Center for Applied Mathematics and Theoretical Physics,\\
University of Maribor, SI20000 Maribor, Slovenia\\ \vspace{.2cm}
${}^4$ Arnold-Sommerfeld-Center for Theoretical Physics,\\
Ludwig-Maximilians-Universit\"at, 80333 M\"unchen, Germany\\ \vspace{.2cm}
${}^5$CERN Theory Department, CH-1211 Geneva, Switzerland}
}

\vspace*{0.8cm}
\end{center}
\vspace*{.5cm}

\noindent
We explore higher-form symmetries of M- and F-theory compactified on elliptic fibrations, determined by the topology of their asymptotic boundaries.
The underlying geometric structures are shown to be equivalent to known characterizations of the gauge group topology in F-theory via Mordell--Weil torsion and string junctions.
We further study dimensional reductions of the 11d Chern--Simons term in the presence of torsional boundary $G_4$-fluxes, which encode background gauge fields of center 1-form symmetries in the lower-dimensional effective gauge theory.
We find contributions that can be interpreted as 't Hooft anomalies involving the 1-form symmetry which originate from a fractionalization of the instanton number of non-Abelian gauge theories in F-/M-theory compactifications to 8d/7d and 6d/5d.

\thispagestyle{empty}
\clearpage

\setcounter{page}{1}


\newpage

\begingroup
  \flushbottom
 \setlength{\parskip}{0pt plus 1fil} 
 \tableofcontents
 \newpage
\endgroup
\newpage

\section{Introduction and Summary}

Geometric engineering provides a powerful framework to study quantum field theories and their non-perturbative aspects.
Exploiting known features of a higher-dimensional theory on spacetime ${\cal M}$, one can uncover details of lower dimensional field and gravitational theories on $M$ by ``engineering'' a compactification ${\cal M} = M_D \times Y_d$ on an internal space $Y_d$.
In this approach, physical data of the $D$-dimensional theory on spacetime $M_D$ are mapped, using a ``dictionary'' specific to the theory on ${\cal M}$, onto properties of the internal space $Y_d$.\footnote{Throughout this work, we denote by $d$ the dimension over $\mathbb{R}$ of the internal space $Y_d$.}
$Y_d$ can then be studied using geometric tools which are not necessarily bound by limitations such as perturbative control.
The success of this process clearly hinges on the ``completeness'' of this dictionary, i.e., our ability to identify the relevant geometric structures associated to a particular physical aspect.

One such aspect is the set of generalized, or higher-form symmetries \cite{Gaiotto:2014kfa} of a quantum field theory.
Formulating their corresponding ``dictionary entries'' in various compactification scenarios in string theory has attracted a lot of recent attention \cite{DelZotto:2015isa,Garcia-Etxebarria:2019cnb,Dierigl:2020myk,Morrison:2020ool,Albertini:2020mdx,Bah:2020uev,Closset:2020scj,DelZotto:2020esg,Bhardwaj:2020phs,DelZotto:2020sop,Bhardwaj:2021pfz,Apruzzi:2021phx,Apruzzi:2021vcu, Hosseini:2021ged}.
In this work, we extend the discussion to compactifications of F- and M-theory on elliptically-fibered Calabi--Yau two- and three-folds $Y_d \stackrel{\pi}{\rightarrow} \mathfrak{B}_{d-2}$ \cite{Vafa:1996xn}.

We will focus on discrete 1-form symmetries $\Gamma$ that arise as the center symmetry of non-Abelian gauge dynamics, and whose gauging enforces the non-trivial gauge group topology $G = G_\text{sc} / \Gamma$, where $G_\text{sc}$ is the simply-connected gauge group with center $Z(G_\text{sc}) \supset \Gamma$ \cite{Gaiotto:2014kfa}.
For $\mathfrak{B}$ a compact base (and hence with gravity dynamical), the latter has a characterization in terms of the Mordell--Weil group of $\pi: Y \rightarrow \mathfrak{B}$ \cite{Aspinwall:1998xj,Mayrhofer:2014opa,Cvetic:2017epq}, or string junctions on $\mathfrak{B}$ \cite{Fukae:1999zs,Guralnik:2001jh}.
Using M-/F-duality, we will establish the explicit connection of these descriptions to the asymptotic $G_4$-fluxes that encode the 1-form symmetries in M-theory compactified on $Y$ \cite{Morrison:2020ool,Albertini:2020mdx}.

We will approach this by inspecting the local geometry defining a non-Abelian gauge algebra $\mathfrak{g}$ associated to a simply-connected group $G_\text{sc}$.
That is, $Y \rightarrow \mathfrak{B}$ is a non-compact fibration, with singular fibers realizing $\mathfrak{g}$ in F-theory.
Reducing the theory on an $S^1$ yields M-theory compactified on $Y$. 
The topology of the asymptotic boundary $\partial Y$ --- which encodes the asymptotic fluxes, and thus the 1-form symmetry in M-theory on $Y$ --- is determined by the $SL(2,\bbZ)$ monodromy around the singular fibers in the bulk of $Y$, which in turn is related to the (local) Mordell--Weil group as well as asymptotic string junctions.
Physically, this identifies the possible line operators charged under the 1-form symmetry from M2-branes wrapping non-compact 2-cycles in M-theory, with asymptotic $(p,q)$-string states in F-theory.
A valid compact model can be understood as ``gluing together'' several local patches along their boundaries.
In general, only a subgroup of the 1-form symmetry in each patch will survive, since new massive states break part of the 1-form symmetry explicitly.
The ``compatible'' boundary fluxes are then exactly captured by the global sections of the glued geometry.
Moreover, the cycles associated to the global sections of the geometry become compact, and one has to sum over the distinct flux configurations representing the modified gauge backgrounds of non-simply connected gauge groups. At the same time, the now dynamical magnetic states break the dual $(D-3)$-form symmetry explicitly.

An interesting aspect of 1-form symmetries is their 't Hooft anomalies.
Specifically, for center 1-form symmetries of gauge theories in spacetime dimension $D > 4$, there is a potential mixed anomaly involving the $(D-5)$-form instanton $U(1)$ symmetry \cite{Apruzzi:2020zot,Cvetic:2020kuw,BenettiGenolini:2020doj}.
This anomaly is a consequence of the fractionalization of the instanton number in the presence of a non-trivial background field for the 1-form center symmetry \cite{Gaiotto:2014kfa,Kapustin:2014gua,Gaiotto:2017yup,Cordova:2019uob}.
In ${\cal N}=1$ compactifications of M-theory to $D=7$ and $D=5$, we show that this anomaly arises from the reduction of the eleven-dimensional (11d) Chern--Simons term in the presence of asymptotic fluxes for $G_4$:
By expressing the boundary contributions as a \emph{fractional} linear combination of compactly supported fluxes, we derive the fractional instanton shift on the Coulomb branch of the effective gauge theory.
We demonstrate that, in case the compactification space $Y$ is elliptically fibered, this anomaly matches that of the six- or eight-dimensional (6d/8d) F-theory compactification \cite{Apruzzi:2020zot,Cvetic:2020kuw}.
Intriguingly, the M-theory computation reveals a mixed anomaly between two 1-form symmetries in 5d, which uplifts to a mixed anomaly between the 6d center symmetry, and a discrete 2-form symmetry of instanton strings \cite{DelZotto:2015isa}.
Moreover, we find for 5d gauge theories with a genuine 5d UV fixed point, that the geometrically determined instanton shift deviates from the value naively expected from the effective gauge description.
This indicates a non-perturbative correction to the 't Hooft anomaly from the superconformal dynamics at the UV fixed point, which would be interesting to scrutinize in the future.
It is important to point out that there can potentially be counterterms, e.g., from topological sectors, cancelling these anomalies field-theoretically, which in M-theory compactifications are not arising from the 11d Chern--Simons term.
We refer to a recent work \cite{Apruzzi:2021vcu} where an example of such a topological sector is discussed in the context of M-theory engineering of 5d SCFTs .

The rest of the paper is organized as follows.
In Section \ref{sec:3app}, we study the higher-form symmetries of M-theory on elliptically fibered Calabi--Yaus in the framework put forward in \cite{Morrison:2020ool,Albertini:2020mdx}.
In Section \ref{sec:geometry}, we then compare the results with known characterizations of the gauge group in F-theory via the Mordell--Weil group \cite{Aspinwall:1998xj,Mayrhofer:2014opa} and string junctions \cite{Fukae:1999zs,Guralnik:2001jh}.
For simplicity, we focus mostly on F-/M-theory compactifications to 8d/7d, where the correspondences between these different approaches can be made concrete.
In Section \ref{sec:anomalies}, we analyze the dimensional reduction of the M-theory Chern--Simons term with boundary fluxes that parametrize the center 1-form symmetry of gauge theories, and derive their 't Hooft anomalies associated with instanton fractionalization for M-/F-theory compactifications to 7d/8d as well as 5d/6d.
Some computational details are collected in the appendices.

\section{Center Symmetries of M-theory on Elliptic Fibrations}
\label{sec:3app}

Yang--Mills theories with a fixed non-Abelian gauge algebra $\mathfrak{g}$ can have different topologies for its gauge group $G$, which generally takes the form
\begin{align}
G = \frac{G_{\text{sc}}}{\mathcal{Z}} \,.
\label{eq:modgauge}
\end{align}
Here, $\mathcal{Z}$ is a subgroup of the center $Z(G_{\text{sc}})$ of the simply-connected group $G_{\text{sc}}$ associated to $\mathfrak{g}$.
In the context of generalized global symmetries \cite{Gaiotto:2014kfa}, the non-trivial global structure \eqref{eq:modgauge} arises from gauging the subgroup ${\cal Z}$ of the global $Z(G_\text{sc})$ 1-form symmetry, which act on electric (Wilson) line charges of $G_\text{sc}$.
The presence of dynamical charged particles in representations ${\bf R}_i$, which in general do not need to be massless, explicitly breaks the center 1-form symmetry to the subgroup of $Z(G_\text{sc})$ that leaves all ${\bf R}_i$ invariant.
This happens due to the fact that the objects charged under the electric 1-form symmetries, i.e., Wilson line operators, can end on these charged particle states, and cease to define 1-form charges.

In $D$ spacetime dimensions, a $\mathfrak{g}$ gauge algebra also has a dual magnetic $Z(G_\text{sc})$ $(D-3)$-form symmetry, which acts on magnetically charged objects.
There is a mixed 't Hooft anomaly between the electric and magnetic higher-form symmetries, which form a so-called ``defect group structure'' \cite{Freed:2006ya,Freed:2006yc,DelZotto:2015isa}, and which is a generalization of Dirac's quantization condition for electric and magnetic charges.
In terms of the defect group, the global form \eqref{eq:modgauge} can also be understood as a choice of ${\cal Z} = \pi_1(G)$ magnetic higher-form symmetry, together with a ``mutually-local'' electric 1-form symmetry.
That is, the electric flux operators present in the $G = G_\text{sc}/{\cal Z}$ theory have integral pairing under the defect group pairing with the magnetic flux operators of the $(D-3)$-form ${\cal Z}$ symmetry.

In string theory realizations of quantum field theories, the charged objects of higher-form symmetries generally arise from branes wrapping asymptotic cycles (more precisely, relative cycles with respect to the asymptotic boundary) of appropriate dimensions in the non-compact internal space $Y$ \cite{DelZotto:2015isa,Garcia-Etxebarria:2019cnb,Morrison:2020ool,Albertini:2020mdx,Bah:2020uev,Closset:2020scj,DelZotto:2020esg,Bhardwaj:2020phs,DelZotto:2020sop,Bhardwaj:2021pfz,Apruzzi:2021phx,Apruzzi:2021vcu}.
These wrapped branes generate flux quanta, whose spacetime part represents the flux operators associated to the charged objects, and whose internal pieces are characterized by cohomology classes on the asymptotic boundary $\partial Y$.
In gauging the electric 1-form ${\cal Z}$ symmetry, the $G_\text{sc}/{\cal Z}$ theories include the magnetically charged objects from branes wrapping the corresponding relative cycles, which transform under the $(D-3)$-form ${\cal Z}$ symmetry.

In gravitational theories, global symmetries, including higher-form symmetries, are believed to be inconsistent \cite{Banks:1988yz,Kallosh:1995hi,Banks:2010zn,Harlow:2018tng}.
Since, in a theory with gauge group $G_\text{sc}/{\cal Z}$, the defect group structure forbids the simultaneous gauging of both ${\cal Z}$ 1-form and $(D-3)$-form symmetries, the magnetic one has to be broken.
Because the compactification space $Y$ is compact in gravitational models, this breaking happens explicitly due to the magnetically charged objects becoming dynamical (as the relative cycles become compact themselves).
As we will highlight below in the M-theory framework (and, by duality, also in F-theory), one can think of the compact model arising from gluing together local (non-compact) patches along their boundaries.

\subsection{Higher-Form Symmetries in M-Theory}
\label{subsec:Mapp}

In M-theory compactifications on local (non-compact) Calabi--Yau manifolds $Y_d$, the information about the 1-form symmetries is encoded in terms of geometric data \cite{Morrison:2020ool, Albertini:2020mdx}:
The electrically charged objects (Wilson lines) $\Gamma_{\text{el.}}$ are associated to M2-branes that stretch from the asymptotic boundary to the interior of $Y_d$, and are classified by classes in the relative homology $H_2 (Y_d, \partial Y_d)$.\footnote{Note that unless otherwise specified, all (co)homology groups $H_n(Y; R) \equiv H_n(Y)$ have coefficients $R = \bbZ$, which is suppressed in the notation.}
However, by an analog of 't Hooft's screening argument, their 1-form symmetry charges are subject to an equivalence relation induced by the addition of M2-branes wrapped over compact 2-cycles.
Mathematically, this is encoded in the quotient
\begin{align}\label{eq:1-form-charges_general}
	\Gamma_\text{el.} \equiv \Gamma = \frac{H_2(Y_d, \partial Y_d)}{\text{im}(\jmath_2)} \cong \frac{H_2(Y_d, \partial Y_d)}{\text{ker}(\partial_2)} \cong \text{im}(\partial_2) \subset H_1(\partial Y_d) \, ,
\end{align}
extracted from the long exact sequence of relative homology,\footnote{The map $\imath$ is induced by the inclusion $\partial Y \hookrightarrow Y$, and $\jmath_n$ is induced by the quotient map onto relative $n$-chains. As usual $\partial_n$ denotes the boundary map.}
\begin{align}\label{eq:long_seq_rel_hom}
\ldots \rightarrow H_n (\partial Y_d) \overset{\imath_n}{\rightarrow} H_n (Y_d) \overset{\jmath_n}{\rightarrow} H_n (Y_d, \partial Y_d) \overset{\partial_n}{\rightarrow} H_{n-1} (\partial Y_d) \rightarrow \ldots \, .
\end{align}

Similarly, by wrapping M5-branes over the relative $(d-2)$-cycles one obtains extended magnetically charged objects in the effective theory:
\begin{align}\label{eq:mag-charges_general}
	\Lambda_\text{mag.} \equiv \Lambda = \frac{H_{d-2}(Y_d, \partial Y_d)}{\text{im}(\jmath_{d-2})} \cong \frac{H_{d-2}(Y_d, \partial Y_d)}{\text{ker}(\partial_{d-2})} \cong \text{im}(\partial_{d-2}) \subset H_{d-3}(\partial Y_d) \, ,
\end{align}
The ``defect group'' pairing between the electric and magnetic charges is then given by the torsion linking pairing
\begin{align}\label{eq:linking_pairing_homology}
	L: \text{Tors}(H_1(\partial Y_d)) \times \text{Tors}(H_{d-3}(\partial Y_d)) \rightarrow \mathbb{Q}/\bbZ \, .
\end{align}
As a generalization of the requirement of mutual locality, imposed by Dirac quantization condition, in 4d, a choice of physical electric charges $\{ \Omega \} \subset \Gamma$ enforces the restriction $L(\Omega, \tilde{\Omega}) = 0$ for allowed magnetic charges $\{\tilde\Omega\} \subset \Lambda$, and vice versa. 

To compute \eqref{eq:1-form-charges_general} and \eqref{eq:mag-charges_general}, we assume that $H_n(Y_d)$ is torsion-free for any $n$ (which holds for all cases relevant to the present discussion).
Then Poincar\'e--Lefschetz duality and the Universal Coefficient theorem provide the identification
\begin{align}
	H_n(Y_d, \partial Y_d) \cong \text{Hom}(H_{d-n}(Y_d),\mathbb{Z}) \, .
\end{align}
Furthermore, $Y_d$ comes with an intersection pairing,
\begin{align}
	\langle \cdot , \cdot \rangle_n : H_{n}(Y_d) \times H_{d-n}(Y_d) \rightarrow \mathbb{Z} \, .
\end{align}
Using the above isomorphism, the maps $\jmath_n$ in \eqref{eq:long_seq_rel_hom} are then given by 
\begin{align}\label{eq:j_map_general}
  \jmath_n(\upsilon) = \langle \upsilon, \cdot \rangle_{n} = \langle \cdot, \upsilon \rangle_{d-n} \in \text{Hom}(H_{d-n}(Y_d), \bbZ) \, .
\end{align}
Picking a basis $\sigma_a$ for $H_{d-2}(Y_d)$ and a basis $\gamma_i$ for $H_2(Y_d)$ defines the $r_{d-2} \times r_2$ intersection matrix $M_{ai} = \langle \sigma_a, \gamma_i \rangle_{d-2} \equiv \langle \sigma_a, \gamma_i \rangle$, where $r_k = \text{rank}(H_k(Y_d))$, with $r_2 - r_{d-2} \equiv f \geq 0$.
Then $\jmath_{d-2}(\sigma_a) = \sum_i M_{ai} \eta_i$, where $\eta_i \in \text{Hom}(H_2(Y_d), \bbZ)$ is the dual basis of $\gamma_i$, i.e., $\eta_i(\gamma_j) = \delta_{ij}$.
Similarly, $\jmath_2(\gamma_i) = \sum_a (M^T)_{ia} \nu_a$, with $\nu_a \in \text{Hom}(H_{d-2}(Y_d),\bbZ)$ dual to $\sigma_a$.
Through a Smith decomposition,
\begin{align}\label{eq:smith_decomp}
	 M_{ai} = \sum_{b, j} \, S_{ab} D_{b j} \, T_{ji} \,,
\end{align}
with $S \, \,(r_{d-2} \times r_{d-2})$, $T \, \, (r_2 \times r_2)$ invertible integer matrices, and
\begin{align}
	  D_{b j} = \begin{pmatrix}
		N_1 & 0 & \ldots & 0 & 0 & \ldots \\
		0 & N_2 & \ldots & 0 & 0 & \ldots \\
		\vdots & \vdots & \ddots & \vdots & \vdots \\
		0 & 0 & \ldots & N_{r_{d-2}} & 0 & \ldots
	\end{pmatrix}_{bj} \, ,
\end{align}
we have
\begin{align}\label{eq:explicit_formula_Gamma}
	\begin{split}
		\Gamma & = \frac{H_2(Y_d, \partial Y_d)}{\text{im}(\jmath_2)} = \frac{\text{Hom}(H_{d-2}(Y_d), \mathbb{Z})}{\text{im}(D^T)} \cong \bigoplus_{k=1}^{r_{d-2}} \mathbb{Z}_{N_k} \, , \\
		\Lambda & = \frac{H_{d-2}(Y_d, \partial Y_d)}{\text{im}(\jmath_{d-2})} = \frac{\text{Hom}(H_{2}(Y_d), \mathbb{Z})}{\text{im}(D)} \cong \Gamma \oplus \mathbb{Z}^f \, .
	\end{split}
\end{align}
More precisely, the Smith decomposition tells us that we can define new bases $\xi_a = \sum_b (S^{-1})_{ab} \, \sigma_b$ for $H_{d-2}(Y_d)$ and $\epsilon_i = \sum_j T_{ij} \, \eta_j$ for $\text{Hom}(H_2(Y_d),\bbZ)$ such that $\jmath_{d-2}(\xi_a) = N_a \, \epsilon_a$ (no sum over $a = 1,...,r_{d-2}$), and similarly for $H_2(Y_d)$ and $\text{Hom}(H_{d-2}(Y_d), \bbZ)$.
For an $N_t$-torsional generator $T \in \Lambda_\text{mag.}$, we thus have a representative $\tilde\sigma_t \in H_{d-2}(Y_d, \partial Y_d)$, satisfying
\begin{align}\label{eq:lin_rel_homology}
  N_t \, \tilde\sigma_t & = \sum_a \lambda_a \, \jmath_{d-2}(\sigma_a) \, , \\
  \text{with} \quad \lambda_a & = (S^{-1})_{ta} \, . \label{eq:lambda_via_smith_decomp}
\end{align}

There is a dual description of the higher-form symmetries in terms of boundary conditions for background fluxes.
In this framework, the defect group structure arises from the non-commutativity of general flux operators at the asymptotic boundary of the compactification space, measured precisely by (the cohomological version of) the linking pairing \eqref{eq:linking_pairing_homology} \cite{Freed:2006ya, Freed:2006yc,DelZotto:2015isa,Garcia-Etxebarria:2019cnb}. In the effective field theory, the boundary conditions for background fluxes in the higher-dimensional theory parametrize allowed background gauge fields $B$ of higher-form symmetries. 
These restrict the possibilities to wrap M2- and M5-branes over elements in relative homology and in turn constrain the spectrum of extended operators, see \cite{Morrison:2020ool}. 
In Section \ref{sec:anomalies}, we will discuss how the background gauge fields enter the dimensional reduction of the M-theory Chern--Simons term.

\subsection{Higher-Form Symmetries on Elliptic Fibrations}
\label{subsec:Mell}

F-theory describes non-perturbative vacua of type IIB string theory, whose space\-time-de\-pen\-dent axio-dilaton field is captured by the complex structure of an auxiliary torus  (see \cite{Taylor:2011wt, Weigand:2018rez, Cvetic:2018bni} for reviews). 
Therefore, F-theory geometries are described by elliptically-fibered $Y_d$, whose base $\mathfrak{B}_{d-2}$ is part of the physical type IIB spacetime:
\begin{equation}
\begin{split}
  T^2 \enspace \hookrightarrow \enspace & Y_d \\
  & \, \downarrow \\
  & \mathfrak{B}_{d-2}
\end{split}
\end{equation}
By duality, F-theory compactified on $Y_d \times S^1$ is equivalent to M-theory compactified on $Y_d$, whose higher-form symmetries we now examine.

For non-compact backgrounds, i.e., where $\mathfrak{B}_{d-2}$ is non-compact, this induces a fibration structure on the asymptotic boundary as well:
\begin{equation}
\begin{split}
  T^2 \enspace \hookrightarrow \enspace & \partial Y_d \\
  & \, \downarrow \\
  & \partial \mathfrak{B}_{d-2}
\end{split}
\end{equation}
In general, the fibration on $\mathfrak{B}_{d-2}$ has singular fibers over (complex) codimension-one loci, which themselves extend to the asymptotic boundary $\partial \mathfrak{B}_{d-2}$.
Their effect on the boundary homology depends very much on the precise type of singular fibers.
It would be important to study such examples in more detail, e.g., in the context of F-/M-theory realizations of 6d/5d SCFTs.

\begin{figure}[ht]
\centering
\includegraphics[width = 0.5 \textwidth]{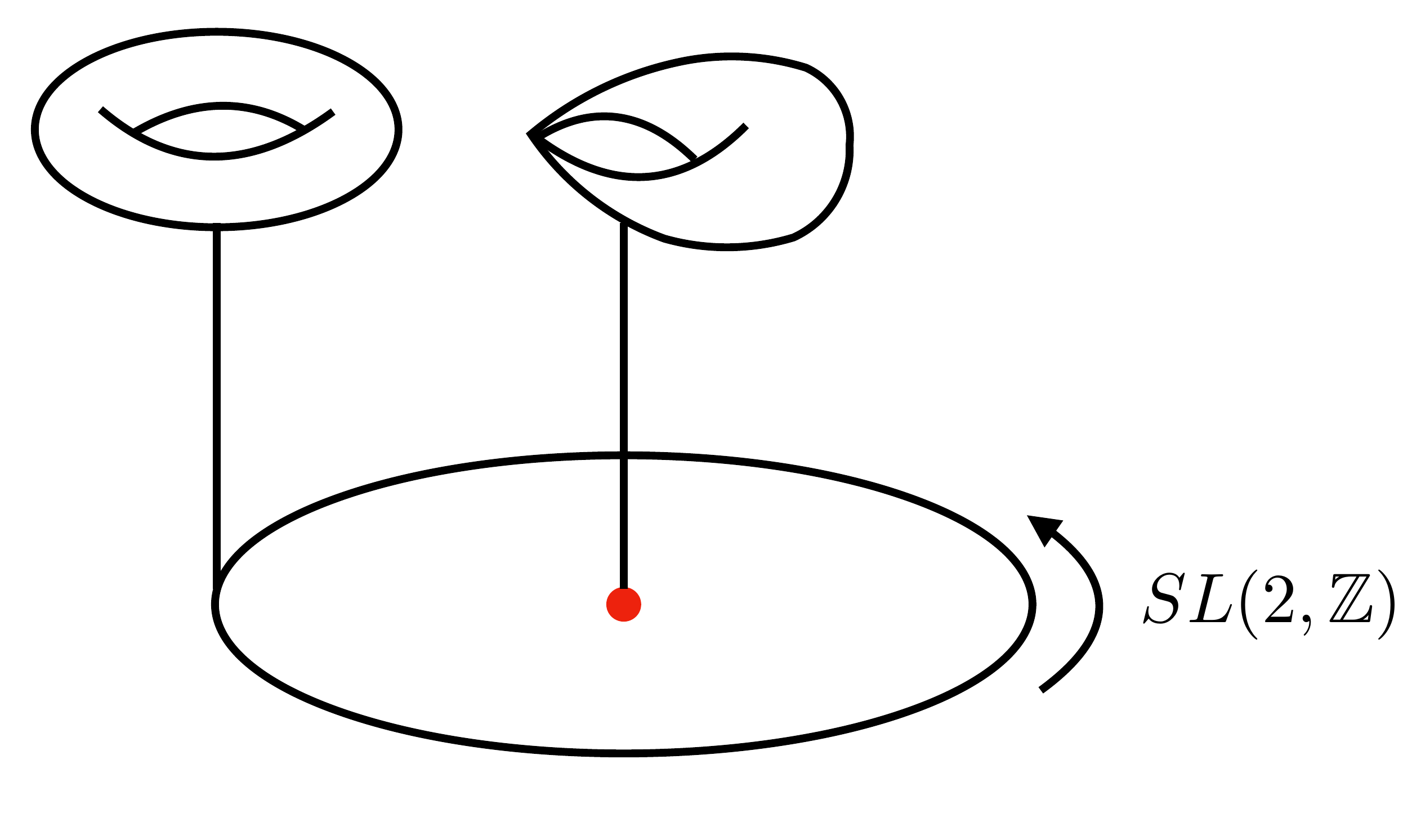}
\caption{Singular geometry with fiber singularity in the interior that induces an $SL(2,\mathbb{Z})$ monodromy around the boundary circle.}
\label{fig:singgeom}
\end{figure}

Aiming for a more intuitive understanding in this work, we avoid these complications, and instead focus on $\dim_\mathbb{R} (Y_d) \equiv d = 4$, i.e., F-/M-theory compactified to 8d/7d.
In this case, the base part of the asymptotic boundary is a circle $\partial \mathfrak{B}_2 \simeq S^1$. Moreover, for situations relevant for supersymmetric F-theory backgrounds,\footnote{That is, backgrounds leading to a local Calabi--Yau 2-fold, i.e., a local patch of an elliptically-fibered K3, for which the base is $\mathbb{P}^1$.} $\mathfrak{B}_2$ itself can be identified with a disc $D_2$. 
We further demand that there is non-trivial gauge dynamics in the effective theory. 
This requires the presence of a singularity in $Y_4$ that can be interpreted as a singular fiber of the elliptic fibration. 
It is the topology of the asymptotic boundary to this fiber singularity that will determine the allowed flux backgrounds and, consequently, the gauge group in the M-theory setup. 
The internal fiber singularity induces a non-trivial fibration on the boundary circle which is associated to a non-trivial $SL(2,\mathbb{Z})$ monodromy, see Figure \ref{fig:singgeom}.\footnote{
The monodromy is not affected by local modifications in the interior such as a resolution of the fiber singularity that corresponds to a Coulomb branch deformation of the effective action derived from M-theory on $Y_4$.}

\subsubsection*{Boundary Geometry}

The boundary\footnote{In the remainder of this section we restrict to the case $d = 4$ and will not denote the dimension of the various spaces explicitly.} $\partial Y$ has the structure of a mapping torus
\begin{align}
\partial Y = T^2 \times [0,1] / \sim \,, \quad \text{with identification} \enspace (x,0) \sim (K(x),1) \,,
\end{align}
where $K: T^2 \rightarrow T^2$ is the overall $SL(2,\bbZ)$ monodromy around all singular fibers in the interior of $\mathfrak{B}_2$. This description allows for an application of a generalized Mayer--Vietoris sequence (see, e.g., \cite{hatcher2002algebraic}), which yields a long exact sequence for the homology groups of $\partial Y$. For our interest the relevant section of this long exact sequence is given by
\begin{equation}\label{eq:mapping_torus_seq}
\begin{split}
0 & \rightarrow H_3 (\partial Y) \rightarrow \underbrace{H_2 (T^2)}_{\cong \bbZ} \stackrel{\alpha}{\rightarrow} \underbrace{H_2 (T^2)}_{\cong \bbZ} \rightarrow H_2 (\partial Y) \rightarrow \underbrace{H_{1} (T^2)}_{\cong \bbZ \oplus \bbZ} \stackrel{\kappa}{\rightarrow} \underbrace{H_{1} (T^2)}_{\cong \bbZ \oplus \bbZ} \rightarrow \\ 
& \rightarrow H_{1} (\partial Y) \rightarrow \underbrace{H_{0} (T^2)}_{\cong \bbZ}  \stackrel{\beta}{\rightarrow} \underbrace{H_{0} (T^2)}_{\cong \bbZ} \rightarrow H_{0} (\partial Y) \rightarrow 0 \,.
\end{split}
\end{equation}
The maps from $H_n (T^2)$ (regarded as a $\bbZ$-module) to itself, denoted by $\alpha$, $\beta$, and $\kappa$, are given by $({\bf 1} - K_{\ast})$, where $K_*$ is the action on $H_n(T^2)$ induced by the $SL(2,\mathbb{Z})$ monodromy $K$.
Since any $SL(2,\bbZ)$ monodromy induces the identify action on $H_2(T^2)$ and $H_0(T^2)$ (it maps the full $T^2$, the generator of $H_2$, onto itself, and one point onto another, both being homologous, i.e., identical in $H_0$), the maps $\alpha$ and $\beta$ are 0.
This fixes $H_3 (\partial Y) \cong \mathbb{Z}$, $H_0 (\partial Y) \cong \mathbb{Z}$, and leaves
\begin{align}\label{eq:split_mapping_torus_seq}
  \begin{split}
    0 \rightarrow \bbZ \rightarrow H_2(\partial Y) \rightarrow \bbZ \oplus \bbZ \stackrel{\kappa}{\rightarrow} \bbZ \oplus \bbZ \rightarrow H_1(\partial Y) \rightarrow \bbZ \rightarrow 0 & \, .
  \end{split}
\end{align}
The remaining homology groups are determined by the monodromy induced map $\kappa$ on $H_1(T^2) \cong \bbZ \oplus \bbZ$, whose generators $(1,0)$ and $(0,1)$ are the usual $\mathcal{A}$- and $\mathcal{B}$-cycle, respectively. The \mbox{(co)kernel} of $\kappa$ splits \eqref{eq:split_mapping_torus_seq} to
\begin{align}
\begin{split}
  0 \rightarrow \bbZ \rightarrow H_2 (\partial Y) \rightarrow \text{ker}(\kappa) \rightarrow 0 \, , \quad \text{and} \quad  0 \rightarrow \text{coker}(\kappa) \rightarrow H_1(\partial Y) \rightarrow \mathbb{Z} \rightarrow 0 \, .
\end{split}
\end{align}
The last term in both of these sequences are free ($\bbZ$ is trivially free, and $\text{ker}(\kappa)$ is a subgroup of a free group $\bbZ \oplus \bbZ$), so both sequences split:
\begin{align}
  H_2(\partial Y) = \bbZ \oplus \text{ker}(\kappa) \, , \quad H_1(\partial Y) = \text{coker}(\kappa) \oplus \bbZ \, .
  \label{eq:boundhom}
\end{align}
From \eqref{eq:mapping_torus_seq}, we see that the $\mathbb{Z}$ factor in $H_2(\partial Y)$ originates from $H_2(T^2)$, i.e., is generated by the class $\mathfrak{f}$ of the torus fiber.
Meanwhile, the $\mathbb{Z}$ factor in $H_1 (\partial Y)$ comes from $H_0(T^2)$, hence corresponds to a marked point on the fiber, i.e., a section of the torus bundle which is a copy of the base circle. We will identify this class with the restriction of the zero-section to the boundary $S_0|_{\partial Y}$.

With a list of fiber singularities provided by the Kodaira classification, see e.g. \cite{Fukae:1999zs}, and their induced $SL(2,\mathbb{Z})$ monodromy up to an overall conjugation, we can determine the respective torsion groups, cf.~Table \ref{tab:fibers_and_kappa}.
\begin{table}[ht]
\renewcommand{\arraystretch}{1.25}
\begin{align*}
\begin{array}{| c | c | c | c | c |}
\hline
\text{fiber type} & \text{brane content} & \mathfrak{g} & \kappa & H_1 (\partial Y, \bbZ)_{\text{tors}} = \text{coker}(\kappa)_\text{tors} \\ \hline \hline
I_N & A^N & \mathfrak{su}(N) & \begin{pmatrix} 0 &  N \\ 0 & 0 \end{pmatrix} & \mathbb{Z}_N \\ \hline
II & A C & - & \begin{pmatrix} 0 & 1 \\ -1 & 1 \end{pmatrix} & - \\ \hline
III & A^2 C & \mathfrak{su}(2) & \begin{pmatrix} 1 & 1 \\ - 1 & 1 \end{pmatrix} & \mathbb{Z}_2 \\ \hline
IV & A^3 C & \mathfrak{su}(3) & \begin{pmatrix} 2 & 1 \\ -1 & 1 \end{pmatrix} & \mathbb{Z}_3 \\ \hline
I_{2n}^* & A^{4 + 2n} B C & \mathfrak{so}(4n + 8) & \begin{pmatrix} 2 & -2n \\ 0 & 2 \end{pmatrix} & \mathbb{Z}_2 \oplus \mathbb{Z}_2 \\ \hline
I_{2n+1}^* & A^{5 + 2n} BC & \mathfrak{so}(4n + 10) & \begin{pmatrix} 2 & - (2n+1) \\ 0 & 2  \end{pmatrix} & \mathbb{Z}_4 \\ \hline
IV^* & A^5 B C^2 & \mathfrak{e}_6 & \begin{pmatrix} 2 & -1 \\ 1 & 1 \end{pmatrix} & \mathbb{Z}_3 \\ \hline
III^* & A^6 B C^2 & \mathfrak{e}_7 & \begin{pmatrix} 1 & -1 \\ 1 & 1 \end{pmatrix} & \mathbb{Z}_2 \\ \hline
II^* & A^7 B C^2 & \mathfrak{e}_8 & \begin{pmatrix} 0 & -1 \\ 1 & 1 \end{pmatrix} & - \\ \hline
\end{array}
\end{align*}
\caption{Simple algebras $\mathfrak{g}$ realized via Kodaira fibers / $[p,q]$-7-branes, and the corresponding homology map $\kappa$, as well as $\text{coker}(\kappa)_{\text{tors}} \cong H_1(\partial Y, \bbZ)_\text{tors}$. \label{tab:fibers_and_kappa}}
\end{table}
This coincides with the table given in \cite{Garcia-Etxebarria:2019cnb, Albertini:2020mdx} for M-theory on lens spaces, which realizes all ADE algebras $\mathfrak{g}$, albeit not in an elliptic fibration (and thus have no F-theory uplift).
Moreover, we also find agreement between the linking pairing on boundary torsion cycles, and the defect group of 7d gauge theories with gauge algebra $\mathfrak{g}$, see Appendix \ref{app:defect_group_elliptic}.
This shows that the 7d theories that descend from an 8d F-theory compactification with simple gauge algebra $\mathfrak{g}$ has the expected electric and magnetic higher-form symmetries.
The only difference is that these 7d theories further contain $U(1)$ global symmetry, the Kaluza--Klein $U(1)$, whose background fluxes / asymptotic charges are captured by $\bbZ \subset H_1(\partial Y)$.

\subsubsection*{Bulk Geometry}

Equipped with the boundary homology groups $H_n(\partial Y)$, we can now examine the long exact sequence \eqref{eq:long_seq_rel_hom}, which encodes the extended charged objects under the higher-form symmetries.
For our investigation the relevant part of the long exact sequence above is given by
\begin{align}
\ldots \rightarrow H_2 (\partial Y) \stackrel{\imath_2}{\rightarrow} H_2 (Y) \stackrel{\jmath_2}{\rightarrow} H_2 (Y, \partial Y) \stackrel{\partial_2}{\rightarrow} H_1 (\partial Y) \stackrel{\imath_1}{\rightarrow} H_1 (Y) \rightarrow \ldots \, ,
\end{align}
Let us focus on the case with a single fiber in the interior of $Y$, corresponding to a simple gauge algebra $\mathfrak{g}$.
Then the resolution of this singularity introduces $\text{rank}(\mathfrak{g})$ compact 2-cycles (divisors in $Y$) $\sigma_a$, which intersect according to the Dynkin diagram of $\mathfrak{g}$.
Together with the generic fiber $\mathfrak{f}$, these form a basis for $H_2 (Y)$.
Since $\mathfrak{f}$ is homologous also to the torus fiber on the boundary, i.e., the factor $\bbZ$ in \eqref{eq:boundhom}, we see that $\mathfrak{f} \in \text{im} (\imath_2)= \text{ker} (\jmath_2)$.
This agrees with \eqref{eq:j_map_general}: the generic fiber $\mathfrak{f}$ on elliptic surfaces satisfies the intersection properties $\langle \mathfrak{f}, \mathfrak{f} \rangle = \langle \mathfrak{f}, \sigma_a \rangle = 0$.
On the other hand, two different resolution divisors cannot have the same intersection numbers with all 2-cycles, so $\jmath_2(\sigma_a) = \jmath_2(\sigma_b)$ if and only if $a = b$.
Therefore, the long exact sequence splits into the piece
\begin{align}
0 \rightarrow \langle \sigma_a \rangle \stackrel{\jmath_2}{\rightarrow} H_2 (Y,\partial Y) \stackrel{\partial_2}{\rightarrow} \underbrace{\mathbb{Z} \oplus \text{coker}(\kappa)_{\text{tors}} \oplus  \text{coker}(\kappa)_{\text{free}}}_{H_1(\partial Y)} \stackrel{\imath_1}{\rightarrow} H_1 (Y) \rightarrow \dots \,.
\end{align}

We have already explained above that the factor $\bbZ \equiv \bbZ_\text{KK} \subset H_1(\partial Y)$ is in ker$(\imath_1)$, as it encodes the background data for the KK $U(1)$ that is universal in M-theory on elliptic fibrations.
Since $H_1 (Y)$ is torsion-free\footnote{This assumption holds also for the elliptically fibered geometries we consider in this work. It would be interesting to study the physics of higher-form symmetries in models with non-trivial torsion in $H_1(Y)$.}, the torsion part $\text{coker}(\kappa)_{\text{tors}}$ cannot be mapped non-trivially into it.
Hence, $\text{coker}(\kappa)_{\text{tors}} \subset \text{ker}(\imath_1) = \text{im}(\partial_2) \subset \Gamma$ in \eqref{eq:1-form-charges_general}.
Therefore, for any $N_t$-torsional element $T \in \text{coker}(\kappa)_{\text{tors}}$ there is a $\tilde{\sigma} \in H_2(Y, \partial Y)$ with $\partial_2(\tilde{\sigma}) = T$.
Then, $N_t \tilde{\sigma} \in \text{ker}(\partial_2) = \text{im}(\jmath_2)$. 
Since $H_2 (Y, \partial Y) \cong \text{Hom}(H_2(Y), \bbZ)$ is also torsion-free, this means there are non-zero integers $\lambda_a$ such that
\begin{align}
\tilde{\sigma} = \frac{1}{N_t} \sum_a \lambda_a \jmath_2(\sigma_a) \in H_2(Y, \partial Y) \,,
\label{eq:reltors}
\end{align}
where the coefficients $\lambda_a$ can be understood modulo $N_t$ since one can always add an integer linear combination of $\jmath(\sigma_a)$ which is in $\text{ker}(\partial_2)$. 
Of course, these are the same coefficients as in \eqref{eq:lambda_via_smith_decomp}, determined via Smith decomposition of the intersection pairing on $Y$.
For example, as we will compute in Section \ref{subsec:7d_anomalies}, the generator of $N$-torsional boundary 1-cycles for $Y$ containing an $I_N$ fiber is represented by
\begin{align}\label{eq:boundary_torsion_rep_I_N}
\begin{split}
  \tilde\sigma & = \tfrac{1}{N} \sum_{a=1}^{N-1} a\, \jmath_2(\sigma_a) \\
  & = \tfrac{1}{N} \sum_{a=1}^{N-1} \underbrace{\tfrac{a-N}{N}}_{(-C^{-1})_{1,a}} \jmath_2(\sigma_a) \mod \text{ker}(\partial_2) \, ,
\end{split}
\end{align}
where $C$ is the Cartan matrix of $SU(N)$.
In Appendix \ref{app:defect_group_elliptic}, where we compute coker$(\kappa)$ for all Kodaira singularities, we see that only $I_N$ fibers have non-trivial coker$(\kappa)_\text{free} \cong \bbZ$, and that it further maps non-trivially under $\imath_1$.
Therefore, coker$(\kappa)_\text{free}$ never contributes to the higher-form symmetries.

In summary, we have seen that the higher-form symmetries \eqref{eq:explicit_formula_Gamma} of M-theory compactified on a local elliptically fibered four-manifold $Y_4 \equiv Y$ are entirely encoded in terms of the monodromy $K \equiv {\bf 1} + \kappa$ around the singular fiber in $Y_4$:
\begin{align}
  \Gamma \cong \text{coker}(\kappa)_\text{tors} \oplus \bbZ_\text{KK} \, .
\end{align}
In Section \ref{sec:geometry}, we will connect this result with ``established'' methods to describe gauge groups with different center symmetries, namely via Mordell--Weil torsion, and string junctions, and show that coker$(\kappa)_\text{tors}$ indeed describes the higher-form symmetries of the F-theory model in one higher dimension.

\subsection{Semi-simple Algebras, Adjoint Higgsing, and Compact Models}

So far, we have only discussed explicit examples with a single Kodaira fiber in $Y$ with monodromy $K$, corresponding to a simple ADE Lie algebra $\mathfrak{g}$.
In these cases, the boundary homology coker$(\kappa) \subset H_1(\partial Y)$ perfectly captures the higher-form symmetry expected from field theory.
However, since $\kappa = {\bf 1} - K$ is an endomorphism on $H_1(T^2) \cong \bbZ^2$, coker$(\kappa)$ can have at most two torsion factors.
This begs the question how this can capture the center symmetries of a semi-simple algebra like, e.g.,  $\mathfrak{g} = \mathfrak{su}(N)^3$, which can be realized by three $I_N$ fibers in $Y$.

The key missing component is dynamical $\mathfrak{u}(1)$ factors which generically arise in the presence of multiple singular fibers.
To see their importance, consider a model $\tilde{Y}$ with one $I_{N-1}$ and one $I_1$ fiber which are mutually local.
That is, in an $SL(2,\bbZ)$ frame where the monodromy of the $I_{N-1}$ fiber is $K_{N-1} = \left(\begin{smallmatrix} 1 & 1-N \\ 0 & 1 \end{smallmatrix} \right)$, the $I_1$ fiber induces $K_1 = \left(\begin{smallmatrix} 1 & -1 \\ 0 & 1 \end{smallmatrix} \right)$.
Therefore, the overall monodromy is $K = K_{N-1} K_1 = \left(\begin{smallmatrix} 1 & -N \\ 0 & 1 \end{smallmatrix} \right)$, with $\text{coker}(\kappa)_\text{tors} = \text{coker}({\bf 1} - K)_\text{tors} = \bbZ_N$.
But F-theory on $\tilde{Y}$ naively has only an $\mathfrak{su}(N-1)$ algebra, whose center cannot possibly accommodate a $\bbZ_N$ 1-form symmetry.
Moreover, $\bbZ_N$ does not have a $\bbZ_{N-1}$ subgroup, so the boundary homology seems to not capture the higher-form symmetries of $\mathfrak{su}(N-1)$ at all.

\begin{figure}[ht]
\centering
\includegraphics[width = 0.8 \textwidth]{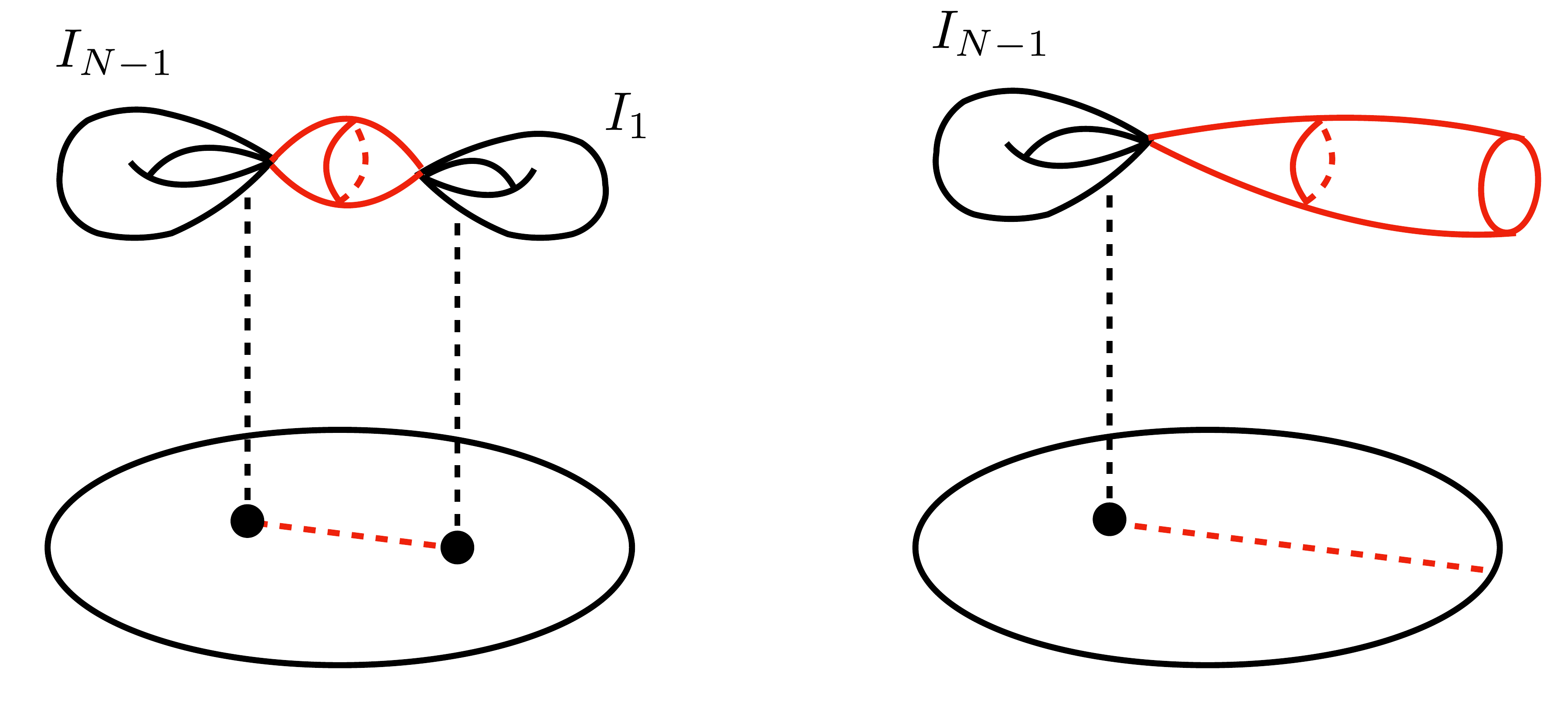}
\caption{Cartoon of a Higgsing transition, where one of the singular fibers is moved outside of the disk, thereby changing the monodromy. \label{fig:1formHiggs}}
\end{figure}

However, there is also the additional $I_1$ fiber.
Because it is mutually local with the $I_{N-1}$ fiber, they share the same the vanishing cycle, in this case the ${\cal A}$-cycle.
By fibering this 1-cycle between the two singular fibers (see left of Figure \ref{fig:1formHiggs}), we obtain a compact 2-cycle $\sigma_N$, in addition to the $N-1$ resolution divisors of the $I_N$ fiber, which gives rise to a dynamical $\mathfrak{u}(1)$.
This $\mathfrak{u}(1)$ can be viewed as the one arising in the adjoint Higgsing $\mathfrak{su}(N) \rightarrow \mathfrak{su}(N-1) \times \mathfrak{u}(1)$, which geometrically precisely corresponds to the deformation of an $I_N$ fiber into an $I_{N-1}$ and a mutually local $I_1$ fiber.

In this Higgsing transition, the fundamental and adjoint representations of the original $\mathfrak{su}(N)$ decompose as
\begin{align}
 {\bf N} \rightarrow ({\bf N-1})_{1} \oplus {\bf 1}_{1-N} \, , \quad  {\bf adj}(N) \rightarrow {\bf adj}(N-1)_0 + ({\bf N-1})_{N} + (\overline{\bf N-1})_{-N} + {\bf 1}_0 \, ,
\end{align}
where the subscripts denote the $\mathfrak{u}(1)$ charge, normalized such that every state has integer charge.
Therefore, the Wilson lines in the fundamental representation of $\mathfrak{su}(N)$, which are the charged objects under the $\mathbb{Z}_N$ 1-form symmetry in the theory prior to Higgsing, gives rise to line operators with $\mathfrak{u}(1)$ charges $1 \mod N$.
However, the bifundamental states from the decomposition of the adjoint representation, which correspond to M2-branes wrapping the red 2-cycle in the left of Figure \ref{fig:1formHiggs}, carry $\mathfrak{u}(1)$ charge $N$.
These screen the $\mathfrak{u}(1)$ charges of line operators in the Higgsed phase, and therefore break the $U(1)$ 1-form symmetry of the $\mathfrak{u}(1)$ gauge factor to $\bbZ_N$.
It is this $\bbZ_N$ 1-form symmetry (and its magnetically dual 3-form symmetry) which is captured by the boundary homology.
The bifundamental states $({\bf N - 1})_N$ also break the $\bbZ_{N-1}$ 1-form symmetry of the $\mathfrak{su}(N-1)$ factor explicitly.
Their presence further forbids the charged objects of the $\bbZ_{N-1}$ magnetic symmetry, but does allow for a linear combination between the $\mathfrak{su}(N-1)$ and the $\mathfrak{u}(1)$ magnetic charges which correspond to M5-branes wrapping the 2-cycle $\sigma_N$.
These have charge $N$ with respect to the magnetic $U(1)$ 3-form symmetry of the $\mathfrak{u}(1)$ gauge factor, so that this is also broken to a $\bbZ_N$.
All this agrees with the fact that the simply-connected group $SU(N)$ actually has $[SU(N-1) \times U(1)]/\bbZ_{N-1}$ as a subgroup, which can be interpreted as the $U(1)$ gauging the center of $SU(N-1)$.

The logic applies to any deformation of Kodaira fibers of ADE type $\mathfrak{g}$ into multiple fibers of type $\mathfrak{g}_i$, corresponding to an adjoint Higgsing which also produces additional $\mathfrak{u}(1)$ gauge factors.
The boundary homology is not affected by such a deformation, and thus the higher-form symmetries of the full system are still those of an $\mathfrak{g}$ gauge theory, albeit embedded as a subgroup of the $U(1)$ higher-form symmetry of the Abelian factors.

To recover the higher-form symmetries of the individual $\mathfrak{g}_i$ factors, one has to decouple the $\mathfrak{u}(1)s$.
Geometrically, this can be easily achieved by pushing all other singular fibers to infinity, or, equivalently, restricting to the local neighborhood of the $\mathfrak{g}_i$ fiber with its monodromy at the boundary.
As an example, consider again the model with an $I_{N-1}$ and an $I_1$ fiber.
As depicted schematically in the right panel of Figure \ref{fig:1formHiggs}, decoupling the $\mathfrak{u}(1)$ sends the $I_1$ fiber to infinity, which turns the previously compact 2-cycle into a relative cycle.
Physically, this turns the dynamical bifundamental states into infinitely massive probe particles in the fundamental representation of $\mathfrak{su}(N-1)$, whose world-lines then constitute the correct charged objects of the $\bbZ_{N-1}$ 1-form symmetry.

Note that not every configuration with multiple fibers allows for an interpretation in terms of a deformation / Higgsing of a single fiber / simple ADE algebra.
In such cases, one has to study more carefully the set of compact 2-cycles stretched between different singular fibers.
For example, if we add an $I_1$ fiber to a mutually non-local $I_{N-1}$ fiber, there is no compact 2-cycle that one can form by fibering a 1-cycle between them, because they have linearly independent vanishing cycles.
Such a configuration would not have an additional $\mathfrak{u}(1)$ factor, and consequently, no way to modify the center symmetries as above.
This can also be seen from the boundary homology.
For concreteness, consider, in an $SL(2,\bbZ)$ frame with $K_{N-1} = \left(\begin{smallmatrix} 1 & N-1 \\ 0 & 1 \end{smallmatrix} \right)$, an $I_1$ fiber with monodromy $K_1 = \left(\begin{smallmatrix} 1 & 0 \\ 1 & 1 \end{smallmatrix} \right)$, such that the overall monodromy is $K = K_{N-1} K_1 = \left(\begin{smallmatrix} N & N-1 \\ 1 & 1 \end{smallmatrix} \right)$.\footnote{The reversed ordering is related to this one by an $SL(2,\bbZ)$ conjugation, so it is equivalent.}
Then, it is straightforward to compute $\text{coker}(\kappa) = \text{coker}\left(\left(\begin{smallmatrix} N-1 & N-1 \\ 1 & 0 \end{smallmatrix} \right) \right) = \bbZ_{N-1}$.

If we have three or more singular fibers, then there can be at most two linearly independent vanishing cycles, simply because any vanishing cycle can be represented as a vector in $\bbZ^2$.
That is, given, e.g., $k$ $I_N$ fibers, each of which has one vanishing cycle ${\cal C}_i$, $i=1,...,k$, there are $(k-2)$ linear relations $\sum_{i=1}^k n^{(\ell)}_i {\cal C}_i = 0$, $\ell = 1,...,k-2$.
If we fiber from the $i$-th singular fiber the vanishing cycle $n^{(\ell)}_i {\cal C}_i$ to a marked smooth fiber $\mathfrak{f}_p$, we have a 2-chain with boundary $n^{(\ell)}_i {\cal C}_i$ in $\mathfrak{f}_p$, which cancel out the boundaries of the corresponding 2-chains from the other singular fibers due to the $\ell$-th relation.
This gives rise to $(k-2)$ compact 2-cycles.
The resulting $\mathfrak{u}(1)$ gauge factors then gauge parts of the overall $(\bbZ_N)^k$ center symmetry, leaving a subgroup with at most two discrete factors.

\subsubsection*{Gluing Patches to Compact Models}

We can use the above insights to describe the process of passing from local to global models.
Effectively, this is done by ``gluing'' the local patches $Y_i$ of individual singular fibers along the boundaries.
In each pairwise gluing, relative 2-cycles in $Y_{i_1}$, whose boundary 1-cycle is a vanishing cycle in $Y_{i_2}$, can be screened by additional compact 2-cycles that are formed between the singular fibers; this corresponds to the situation in Figure \ref{fig:1formHiggs}, read from right to left.
This modifies the higher-form symmetry charges, as seen from the boundary homology in terms of (torsional) 1-cycles on $\partial Y$.

To obtain a compact geometry, we must demand that any 1-cycle in the torus fiber can shrink (possibly after decomposing) on singular fibers, in accordance with the fact that there is no (non-trivial) boundary.
Equivalently, this means that the overall monodromy around all singular fibers must be trivial.
Additionally, for a valid F-theory geometry, $Y_4$ must be a K3-surface, which further limits the possible combination of singular fibers.
A more subtle effect that becomes relevant in compact models is that there might be certain linear relations between 2-cycles, such that the physically distinct number of $\mathfrak{u}(1)$s can be reduced. 
Regardless, compact 2-cycles stretched between several singular fibers gauges a diagonal subgroup of the corresponding center 1-form symmetry, as in the $\mathfrak{su}(N-1) \times \mathfrak{u}(1)$ example above. 
Phrased in the language of string junctions, whose local picture we will discuss momentarily, these phenomena have been discussed in \cite{Guralnik:2001jh}.

For example, a valid K3 can be obtained by gluing together four local patches with an $I_0^*$ fiber, each with monodromy $K = \left( \begin{smallmatrix} -1 & 0 \\ 0 & -1 \end{smallmatrix} \right)$.
Since each $I_0^*$ fiber has two independent vanishing cycles, which also generate the two $\bbZ_2$ factors of $Z(Spin(8))$, one would find $8-2 = 6$ linear relations between them, corresponding to six compact 2-cycles stretching between the four singular fibers.
However, from the compactness condition there are two additional relations among these, such that there are only four independent $\mathfrak{u}(1)$ gauge factors, which in total gives a rank 20 gauge group.\footnote{Note that two of these $\mathfrak{u}(1)$s come from the 8d ${\cal N}=1$ gravity multiplet.
These are always present, though the embedding of the center from the non-Abelian gauge symmetry into them is model specific.
}
Nevertheless the six compact 2-cycles lead to the gauging of six independent $\bbZ_2$ subgroups of the full $(\bbZ_2)^8$ 1-form symmetry \cite{Guralnik:2001jh}.
Two are orthogonal to the $\mathfrak{u}(1)$s, so that the non-Abelian part of the gauge group is $G = Spin(8)^4 / [\bbZ_2 \times \bbZ_2]$, where the denominator is the ``diagonal'' $\bbZ_2 \times \bbZ_2$ subgroup of $(\bbZ_2)^8$.
As for the Abelian factors, one can choose an appropriate basis for them, such that the $\bbZ^{\text{diagonal}}_2 \subset \bbZ_2 \times \bbZ_2$ subgroup of each $Spin(8)$ factor is embedded into one of the $U(1)$s.
The full gauge group is therefore
\begin{align}
  \frac{(Spin(8)^4 / [\bbZ_2 \times \bbZ_2]) \times U(1)^4}{(\bbZ_2)^4} \cong \frac{Spin(8)^4 \times U(1)^4}{(\bbZ_2)^6} \, .
\end{align}

\section{Center Symmetries in M-/F-theory Duality}
\label{sec:geometry}

In global, compact F-theory models, there are two equivalent ways to characterize the gauge group topology: either via the Mordell--Weil group of rational sections \cite{Aspinwall:1998xj,Mayrhofer:2014opa,Cvetic:2017epq} (see also \cite{Grimm:2015wda}), or through so-called (fractional) null junctions \cite{Fukae:1999zs,Guralnik:2001jh}.
The purpose of this section is to relate these ideas to the characterization of the gauge group via higher-form symmetries presented above.
We begin with the string junctions, since these have direct visualizations in terms of relative cycles in the local setting.

\subsection{String Junctions}
\label{subsec:StrJapp}

\begin{figure}[ht]
\centering
\includegraphics[width = .8 \textwidth]{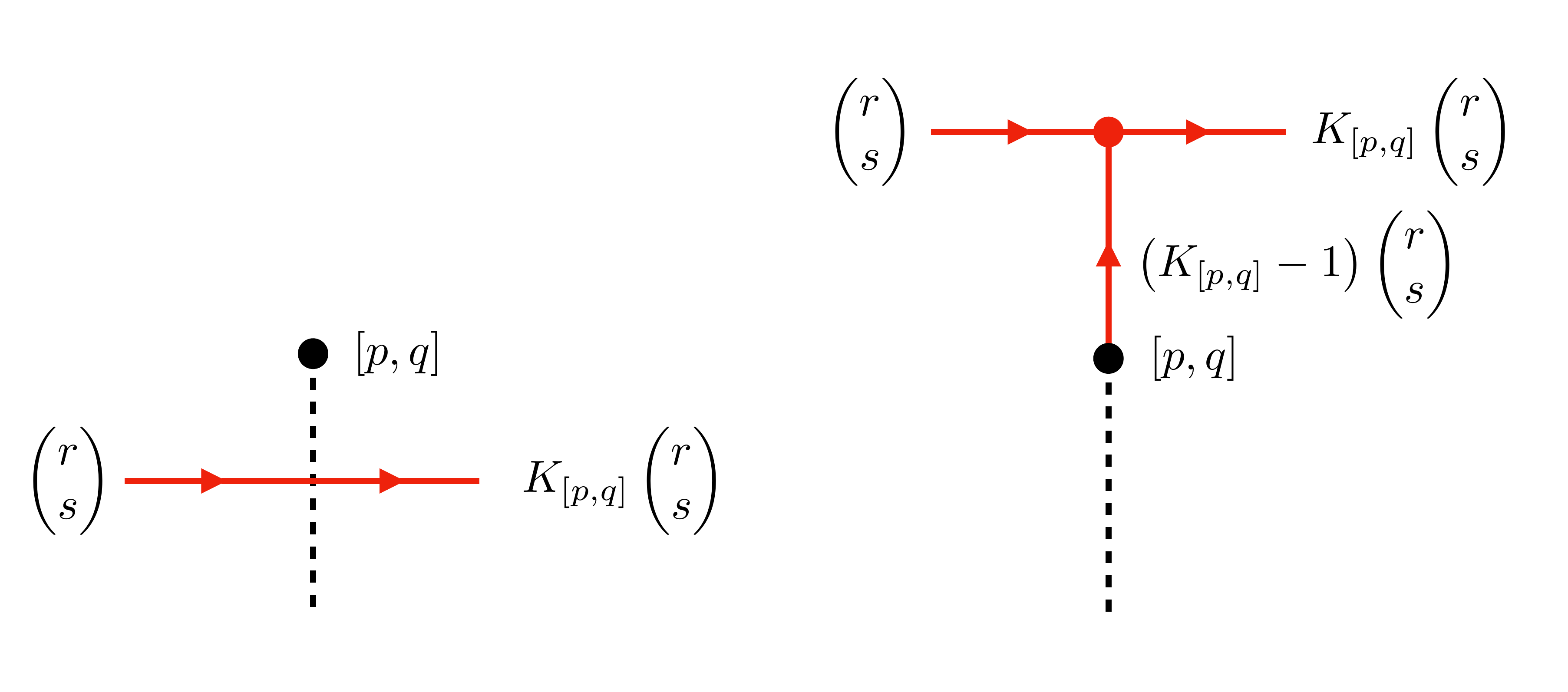}
\caption{Transformation properties of a general $(r,s)$-string passing through the branch-cut of a $[p,q]$-brane.}
\label{fig:junctiondeformation}
\end{figure}

We begin with a brief review of F-theory / type IIB in terms of string junctions.
In this picture, the gauge dynamics are associated to the world-volume of non-perturbative 7-branes, which are classified by their $[p,q]$-type. 
The gauge degrees of freedom on such branes are described by $(p,q)$-strings, i.e., bound states of $p$ fundamental strings and $q$ D1-strings, which can end on a 7-brane of type $[p,q]$ \cite{Gaberdiel:1997ud, Gaberdiel:1998mv}. 
Much like the geometrization of the axio-dilaton in terms of an auxiliary torus, the $(p,q)$-labels encode the transformation properties, or charges, under the $SL(2,\bbZ)$ duality group of 7-branes and strings in type IIB string theory.
The induced $SL(2,\mathbb{Z})$ monodromy around a general $[p,q]$ 7-brane is given by
\begin{align}
K_{[p,q]} = \begin{pmatrix} 1 + p q & -p^2 \\ q^2 & 1 - p q \end{pmatrix} \,.
\label{eq:branemon}
\end{align}
This monodromy transformation is implemented by a branch-cut that emanates from the brane stack and stretches to infinity. When an $(r,s)$-string stretches across this branch-cut its charges change as indicated in Figure \ref{fig:junctiondeformation}.
This configuration can be deformed across the brane. As in Hanany--Witten transitions, a new string connected to the brane stack appears, see Figure \ref{fig:junctiondeformation}.

In F-theory a single $[p,q]$-brane corresponds to an $I_1$ fiber singularity. More general fiber degenerations can then be understood by stacking several 7-branes on top of each other. In the process some of the strings stretching between the individual branes become massless and constitute the gauge theory degrees of freedom on the eight-dimensional brane worldvolume. The overall $SL(2,\bbZ)$ monodromy is given by the product of the constituents. In this way one can reconstruct the full list of Kodaira singularities \cite{Gaberdiel:1997ud,DeWolfe:1998zf}. This determines the gauge algebra of the system, but is not enough to provide information about the gauge group. 
In the following we will focus on the analysis of compactifications to eight dimensions, where 7-branes are parallel, and in which case global tadpole cancellation requires 24 7-branes with overall trivial monodromy, whose transverse positions is parametrized by a $\bbP^1 \cong \mathfrak{B}_2$ that is the base of an elliptically fibered K3-surface.

\begin{figure}[ht]
\centering
\includegraphics[width = 0.9 \textwidth]{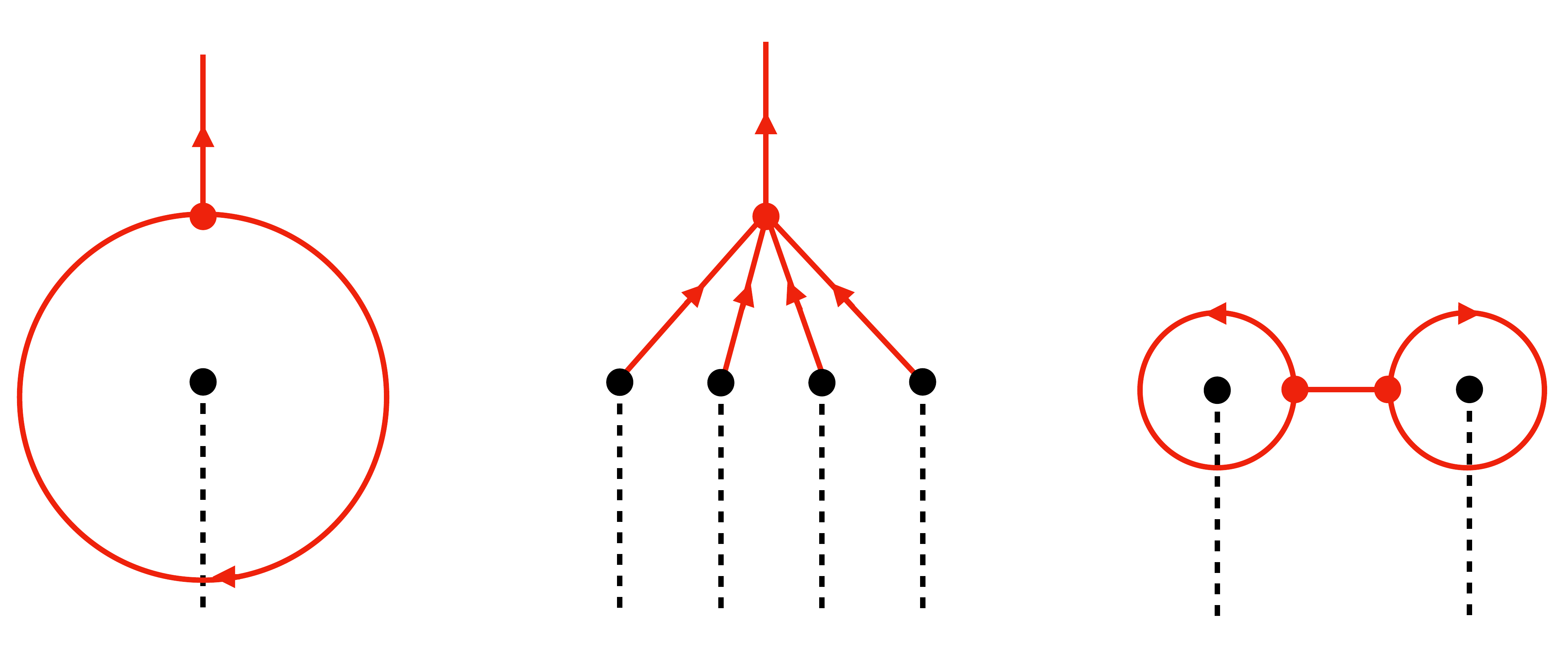}
\caption{A local null junction (left) obtained from encircling a brane stack with a string. In general, it carries an asymptotic charge.
Via a Hanany--Witten transition, the null junction can also be presented as joining prongs from the constituent branes of the stack (middle).
One can connect local null junctions via their asymptotic charges; the global null junctions on a compact $\bbP^1$ have no net asymptotic charge (schematically on the right).
}
\label{fig:nulljunc}
\end{figure}

To gain access to the information of the gauge group one has to analyze the full lattice of string junctions. 
Since Gauss' law forbids the presence of asymptotic charges on \emph{compact} spaces, all junctions are either closed, or have prongs ending on 7-brane stacks.
In other words, no junction can be allowed to have a free prong, whose $(p,q)$ label would be so-called asymptotic charges of the junction.
In determining the gauge group, a special role is played by the so-called null junctions. 
They are constructed by encircling all singularities of the compact model, and thus experience no net monodromy.
These junctions have vanishing pairing\footnote{A precise definition of the junction pairing is given in \cite{DeWolfe:1998zf}. For the present discussion, it suffices to note that this pairing can be identified with the intersection pairing in homology in the dual M-theory frame.} with all other junctions, and can be viewed as a ``trivial'' physical state.
More precisely, since the null junction encircles all 7-branes, one can close the loop ``on the other side'' of the $\bbP^1$, thus removing the string completely.
On the other hand, using the Hanany--Witten transition discussed above one can pull the string through all of the 7-brane stacks, leading to a configuration of a multi-pronged strings of vanishing asymptotic charge that ends on the individual stacks. 
It is obvious that a global null junction can be thought of as the sum of ``local'' null junctions, i.e., junctions that encircle one brane stack, and have a prong that emanates from the circle to connect with other local null junctions.
See Figure \ref{fig:nulljunc} for a schematic depiction of local null junctions.

Zooming onto the local patch around a single brane stack, realizing the algebra $\mathfrak{g}$ with simply-connected cover $G_\text{sc}$, such local null junctions generally carry asymptotic $(p,q)_\text{asymp}$ charge, represented by the prong going off to infinity on the left of Figure \ref{fig:nulljunc}.
If the stack is encircled by a $(r,s)$-string, this charge is
\begin{align}\label{eq:asymp_charge_null_junc}
\begin{pmatrix} p \\ q \end{pmatrix}_{\text{asymp}} = (K - \mathbf{1}) \begin{pmatrix} r \\ s \end{pmatrix} \equiv -\kappa \begin{pmatrix} r \\ s \end{pmatrix} \,,
\end{align}
where $K$ is the monodromy of the stack.
It turns out to be useful to consider \emph{all} possible charges $(r,s)$ such that the asymptotic charges $(p,q)_\text{asymp}$ are integers.
For integral $(r,s)$, the resulting null junction is called a proper, or integer null junction\footnote{Such a junction can be represented by the left panel of Figure \ref{fig:nulljunc}; by making the circle infinitely large, it is obvious that integral null junctions decouple from the local gauge dynamics of the brane stack.
}, whose asymptotic charges are
\begin{align}\label{eq:integer_null_junctions}
  \text{integer null junctions} = \left\{ \left. \kappa \begin{pmatrix} r \\ s \end{pmatrix} \, \right| \, r,s \in \bbZ \right\} = \text{im}(\kappa: \bbZ^2 \rightarrow \bbZ^2) \, .
\end{align}
However, with suitably fractional $(r,s)$, one can generate all integer asymptotic charges $(p,q)_\text{asymp}$, if $\mathfrak{g} \neq \mathfrak{su}(N)$; if $\mathfrak{g} = \mathfrak{su}(N)$, then $\eqref{eq:asymp_charge_null_junc}$ generates all integer charges of the form $(p,0)_\text{asymp}$, up to $SL(2,\bbZ)$ conjugacy.
If one performs a Hanany--Witten transition for these so-called fractional null junctions, then the individual prongs on the constituent branes of the stack (as depicted schematically on the right in Figure \ref{fig:nulljunc}) are in general fractional as well.
Clearly, there is always an integer which multiplies a fractional null junction into an integer null junction.
Then, considering the quotient, we find (fractional null junctions)/(integer null junctions) $\cong \text{coker}(\kappa)_\text{tors}$.

For the various brane configurations that realize Kodaira fibers, we list the generators of local fractional null junctions (also known as \emph{extended weight junctions} \cite{DeWolfe:1998zf}), as well as their fractional prongs ending on the brane constituents of the central brane stack, in \eqref{eq:frac_null_junc_list} \cite{DeWolfe:1998zf,Guralnik:2001jh}.\footnote{We denote the fractional prongs ending on the individual constituents according to their ordering in the second column.}
As is evident from this table, we can identify (fractional null junctions)/(integer null junctions) $\cong Z(G_\text{sc})$.
\begin{table}[ht]
\begin{align}\label{eq:frac_null_junc_list}
\renewcommand{\arraystretch}{1.25}
\begin{array}{| c | c | c |}
\hline
\text{Kodaira / $\mathfrak{g}$} & \text{branes} & \text{generating fractional junction}_{(p,q)_{\text{asymp}}} \\ \hline \hline
I_N \, /\, \mathfrak{su}(N) & A^N & \big\{ \tfrac{1}{N}, \dots, \tfrac{1}{N} \big\}_{(1,0)} \\ \hline
II \, / \, -& A C & - \\ \hline
III \, / \, \mathfrak{su}(2) & A^2 C & \big\{ \tfrac{1}{2}, \tfrac{1}{2}, 0 \big\}_{(1,0)} \, , \quad \big\{\tfrac12, \tfrac12, -1 \big\}_{(0,1)} \\ \hline
IV \, / \, \mathfrak{su}(3) & A^3 C & \big\{ \tfrac{1}{3}, \tfrac{1}{3}, \tfrac{1}{3}, 0 \big\}_{(1,0)} \, , \quad \big\{ \tfrac13,\tfrac13,\tfrac13,-1 \big\}_{(0,1)} \\ \hline
\multirow{2}{*}{$I^*_{n-4} \, / \, \mathfrak{so}(2n)$} & \multirow{2}{*}{$A^{4 + 2n} B C$} & \{ 0,...,0, \tfrac12, \tfrac12 \}_{(1,0)} \\ 
& & \{ \tfrac12,...,\tfrac12, \tfrac{-n-1}{2}, \tfrac{1-n}{2} \}_{(0,1)} \\\hline
\multirow{2}{*}{$IV^* \, / \, \mathfrak{e}_6$} & \multirow{2}{*}{$A^5 B C^2$} & \{ -\tfrac13, ..., -\tfrac13, \tfrac43 , \tfrac23 ,\tfrac23\}_{(1,0)} \\ 
& &   \{ 1, ..., 1, -3,-1,-1\}_{(0,1)} \\ \hline
\multirow{2}{*}{$III^* \, / \, \mathfrak{e}_7$} & \multirow{2}{*}{$A^6 B C^2$} & \{ -\tfrac12, ..., -\tfrac12, 2,1,1\}_{(1,0)} \\ 
& &   \{ \tfrac32, ..., \tfrac32, -5,-2,-2\}_{(0,1)} \\ \hline
\multirow{2}{*}{$II^* \, / \, \mathfrak{e}_8$} & \multirow{2}{*}{$A^7 B C^2$} & \{ -1,...,-1,4,2,2\}_{(1,0)}\\
& & \{ 3,...,3 -11, -5, -5\}_{(0,1)} \\ \hline
\end{array}
\end{align}
\end{table}

In a global model, the requirement of vanishing asymptotic charge ``selects'' a subgroup of the center $Z(G^{(i)}_\text{sc})$ of the $i$-th brane stack, represented by a local fractional null junction, which is then combined with the fractional junctions of other stacks into a global fractional null junction (cf.~right panel of Figure \ref{fig:nulljunc}).
The gauge group $G_\text{global}$ of the full theory then satisfies $\pi_1(G_\text{global}) \cong$(global fractional null junctions)/(global integer null junctions) \cite{Guralnik:2001jh}.

It is important to point out that the null junctions are not actually physical junctions \cite{DeWolfe:1998zf}, as, by construction, they have vanishing pairing with all other junctions.\footnote{Note that the local fractional null junctions have vanishing intersection with all root junctions $\{\sigma_a\}$ in the interior, hence, they can be thought of as trivial linear maps, $\{\sigma_a\} \rightarrow \bbZ$, induced by the junction pairing.}
On the other hand, the asymptotic charge \eqref{eq:asymp_charge_null_junc} is a physical quantity.
Hence, the triviality of the fractional null junction implies that this asymptotic charge, which is carried by the prong stretching to infinity in Figure \ref{fig:nulljunc}, must be generated by the prongs emanating from the individual branes.
The latter can be expressed in terms of a fractional linear combination of the root junctions, as we will demonstrate now.

\begin{figure}[ht]
\centering
\includegraphics[width = 0.6 \textwidth]{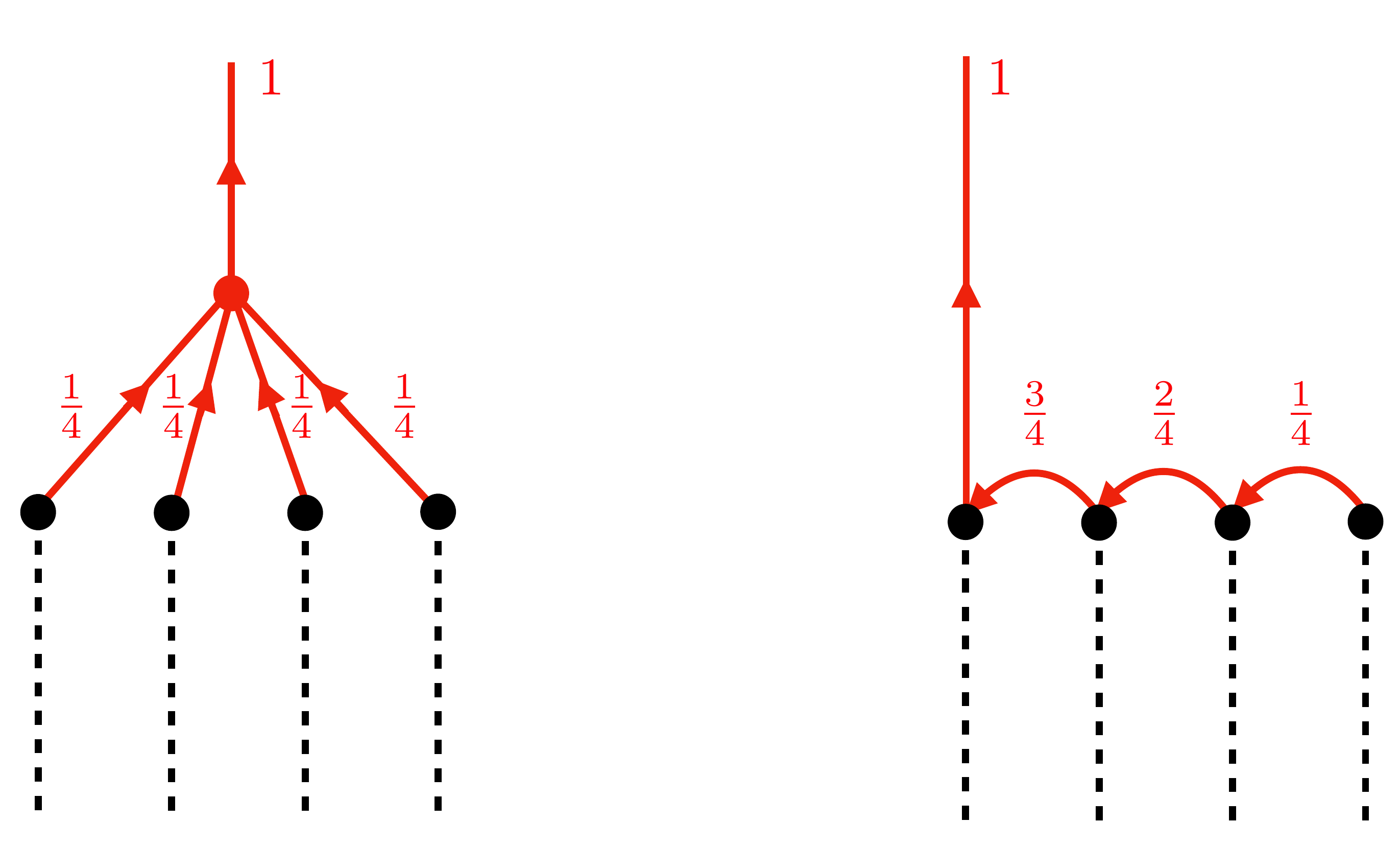}
\caption{Schematic depiction of a local contribution to a fractional null junction in terms of a physical asymptotic junction and a fractional combination of root junctions for $A^4$ stack (which we depicted as separated for convenience).}
\label{fig:nullroots}
\end{figure}

Let us consider a concrete example of a model with $\mathfrak{su}(N)$ gauge algebra.
This is realized by a stack of $N$ mutually local branes, whose $N-1$ simple roots $\sigma_a$ are single-prong strings which stretch between two consecutive branes.
In the notation of \eqref{eq:frac_null_junc_list}, these are
\begin{align}
  \sigma_a = \{0,..., 0, \underbrace{1}_{a\text{-th}}, -1, 0, ..., 0\} \, .
\end{align}
Then, the fractional junction with asymptotic charge $(p,q)_\text{asymp} = (1,0)$ is
\begin{align}
\begin{split}
\big\{ \tfrac{1}{N}, \dots, \tfrac{1}{N} \big\} & = \{1,0, ..., 0\} + \tfrac{1-N}{N} \{1,-1,0,...\} + \tfrac{2-N}{N} \{0,1,-1,0,...\} + ... + \tfrac{1}{N} \{0,...,1,-1\} \\
& = \{1,0,...,0\} + \sum_{a=1}^{N-1} \tfrac{a-N}{N} \sigma_a = \{1,0,...,0\} + \sum_{a=1}^{N-1} (-C^{-1})_{1,a} \, \sigma_a \, ,
\end{split}
\end{align}
where $C$ is the Cartan matrix of $SU(N)$.
This shows the equivalence between different presentations of the asymptotic junction $\big\{ \tfrac{1}{N}, \dots, \tfrac{1}{N} \big\}$, as depicted in Figure \ref{fig:nullroots} for $N=4$.

Because the null junction itself is trivial, we see that the integral prongs stretching to infinity, and which carry the asymptotic charge, are a fractional linear combination of the root junctions:
\begin{align}\label{eq:asymp_junction_suN_generator}
  \{1,0,...,0\} \simeq \sum_{a=1}^{N-1} (C^{-1})_{1,a} \sigma_a \, ,
\end{align}
which, up to a sign, takes the same form as the representation \eqref{eq:boundary_torsion_rep_I_N} in terms of the relative homology of the elliptic fibration.
Moreover, we claimed that the fractional null junctions, modulo integer null junctions, represent the center $\bbZ_N$.
That is, $k \in \bbZ_N$ is represented by the fractional null junction $\big\{ \tfrac{k}{N}, ..., \tfrac{k}{N} \big\}$.
To rewrite this in a non-trivial manner, note that the inverse Cartan matrix of $SU(N)$ is \cite{2017arXiv171101294W}
\begin{align}
  (C^{-1})_{ab} = \text{min}(a,b) - \frac{a\,b}{N} \, ,
\end{align}
which satisfies
\begin{align}
\begin{split}
  k \, (C^{-1})_{1,b} & = k \, \frac{b}{N} \mod \bbZ \\
  & = (C^{-1})_{k b} \mod \bbZ \, .
\end{split}
\end{align}
Then, we have
\begin{align}\label{eq:k_frac_junc}
\begin{split}
  \{k,0,...,0\} & \simeq  \sum_{a=1}^{N-1} k (C^{-1})_{1,a} \, \sigma_a \\
  & = \sum_{a=1}^{N-1} (C^{-1})_{ka} \sigma_a + (\text{root junctions}) \, .
\end{split}
\end{align}

\subsubsection*{1-form Symmetry Charges from String Junctions}

The correspondence between the junction picture and the M-theory description of Section \ref{sec:3app} is on the nose, if we identify string junctions with wrapped M2-branes \cite{Grassi:2013kha}:
a $(p,q)$-prong of a string junction corresponds, in the dual M-theory frame, to M2-branes wrapping a 2-cycle that is the fibration of the 1-cycle
\begin{align}
\mathcal{C} = p \mathcal{A} + q \mathcal{B}
\end{align} 
in the torus fiber over a curve in the base $\mathfrak{B}$.
The splitting of this prong into other prongs $(p_i, q_i)$ corresponds to a linear relation ${\cal C} = \sum_i p_i {\cal A} + q_i {\cal B}$ in $H_1(T^2)$.
The prong can also end on a 7-brane, in which case the cycle ${\cal C}$ is a vanishing cycle in the singular fiber corresponding to the 7-brane.

If a junction has only prongs ending on 7-branes, then these correspond to M2-branes wrapping compact 2-cycles that stretch between fiber singularities in $Y$, i.e., these define elements in $H_2 (Y)$.
In a local model, we have the additional option for a prong to extend to infinity, and thus leaving behind an asymptotic charge $(p,q)_\text{asymp}$.
In the geometry, this then corresponds to a relative cycles in $H_2(Y,\partial Y)$ whose boundary is given by $\mathcal{C} = p{\cal A} + q{\cal B} \subset \partial Y$. 
\begin{figure}
\centering
\includegraphics[width = 0.8 \textwidth]{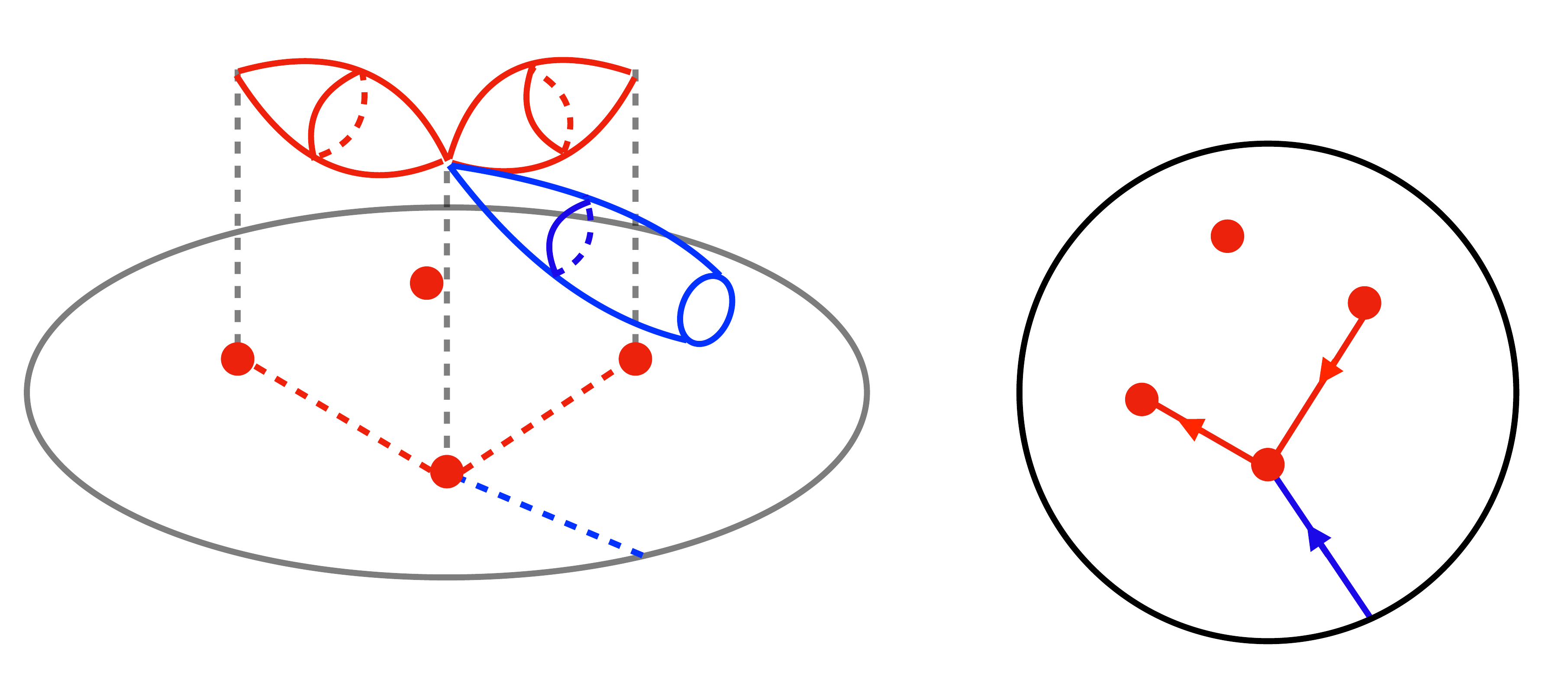}
\caption{The schematic relation, for an $I_4$ singularity, between string junctions and M2-brane states that correspond to roots (red) and asymptotic charges (blue).}
\label{fig:geomjunc}
\end{figure}
Similarly, $\sigma_a \in H_2(Y)$ would have vanishing asymptotic $(p,q)$ charge, and correspond to the root junctions.
See Figure \ref{fig:geomjunc} for a schematic depiction of both types in case of an $I_4$ singularity.\footnote{Note that we are making use here of the equivalence on elliptic surfaces between blow-up resolution, where one introduces 2-cycles into the fiber over a point, and deformation smoothing, where the new 2-cycles arise as fibrations of 1-cycles between the $I_1$ fibers into which a general Kodaira singularity has been deformed into.}

We can now easily identify the geometric counterpart of the local null junctions.
The key is that the asymptotic charges \eqref{eq:asymp_charge_null_junc} of null junction are defined by the same monodromy-induced map $\kappa$ that determines the boundary homology via \eqref{eq:split_mapping_torus_seq}.
Since the set of possible asymptotic $(p,q)$ charges correspond to 1-cycles ${\cal C} = p{\cal A} + q{\cal B}$ on the boundary torus fiber, we immediately see that $\text{coker}(\kappa)_\text{tors} \cong \text{(fractional null junctions)/(integer null junctions)}$.
These boundary 1-cycles are represented in the bulk by the relative 2-cycles \eqref{eq:reltors}, which mirrors the representation of the asymptotic junctions in terms of fractional root junctions in \eqref{eq:asymp_junction_suN_generator} and \eqref{eq:k_frac_junc}.

Without specifying the exact brane content that is encircled by the null junctions, imposing the existence of specific fractional null junctions can restrict the possible $SL(2,\mathbb{Z})$ monodromy induced by the stack. 
Suppose that one demands the existence of fractional null junctions of the form
\begin{align}
\begin{pmatrix} r \\ s \end{pmatrix} = \begin{pmatrix} \tfrac{1}{N} \\ 0 \end{pmatrix} \,.
\label{eq:specnnull}
\end{align}
The requirement of integer asymptotic charge around a stack with monodromy $K = \left(\begin{smallmatrix} a & b \\ c & d \end{smallmatrix} \right) \in SL(2,\mathbb{Z})$ then reads
\begin{align}
(K - \mathbf{1}) \begin{pmatrix} \tfrac{1}{N} \\ 0 \end{pmatrix} = \frac{1}{N} \begin{pmatrix} a-1 \\ c\end{pmatrix} \,,
\end{align}
leading to the constraints $a = 1 \, \text{mod} \, N$, $c = 0 \, \text{mod} \, N$. 
Similarly, one can consider the fractional null junction with \eqref{eq:specnnull}, which encircles the brane stack, and passes its the branch-cut in the opposite direction. 
This corresponds to turning the arrow on the left-hand side of Figure \ref{fig:nulljunc} around. The associated monodromy is generated by $K^{-1}$ and one has
\begin{align}
(K^{-1} - \mathbf{1}) \begin{pmatrix} \tfrac{1}{N} \\ 0 \end{pmatrix} = \frac{1}{N} \begin{pmatrix} d-1 \\ -c \end{pmatrix} \,,
\end{align}
which yields $d = 1 \, \text{mod} \, N$. Together with the constraints above this can be summarized as
\begin{align}
\begin{pmatrix} a & b \\ c & d \end{pmatrix} = \begin{pmatrix} 1 & \ast \\ 0 & 1 \end{pmatrix} \text{ mod } N \,.
\end{align}
This means that the allowed monodromies $K$ are in the congruence subgroup $\Gamma_1 (N)$ of $SL(2,\mathbb{Z})$.\footnote{
In compact models, there are interesting implications for the allowed congruence subgroups \cite{Dierigl:2020lai} imposed by the cobordism conjecture \cite{McNamara:2019rup,Montero:2020icj}.}
An elliptic fibration $Y \stackrel{\pi}{\rightarrow} \mathfrak{B}$ with such a restricted monodromy is known to preserve $N$-torsional points in the fiber \cite{diamond2006first}, which form torsional sections of $\pi$ that are also known to characterize the gauge group topology in F-theory.

\subsection{F-theory and Torsional Sections}
\label{subsec:Fapp}

Sections of an elliptic fibration $\pi: Y_d \rightarrow \mathfrak{B}$ form the so-called Mordell--Weil group, with the zero-section $S_0$ being the neutral element.
It is a finitely generated Abelian group,
\begin{align}
  \text{MW}(\pi) = \mathbb{Z}^s \times \mathbb{Z}_{N_1} \times \mathbb{Z}_{N_2} \, ,
\end{align}
whose torsional part can only contain up to two independent generators, whose orders $N_t$ are bounded by $8$ on compact elliptic fibrations suitable for F-theory models \cite{Hajouji:2019vxs} (see \cite{Park:2011wv,Lee:2019skh,Grassi:2021wii} for discussions on bounds for $s$ in this context).
In compact models, the gauge group is shown to be \cite{Aspinwall:1998xj,Mayrhofer:2014opa,Cvetic:2017epq}
\begin{align}
G = \Big( \prod_i G_{\text{sc},i} \times U(1)^{s} \Big) / (\bbZ_{n_1} \times \dots \times \bbZ_{n_{s}} \times \bbZ_{N_1} \times \bbZ_{N_2}) \, ,
\label{eq:gaugemod}
\end{align}
where $G_{\text{sc},i}$ are the simply-connected non-Abelian groups associated with the gauge algebras $\mathfrak{g}_i$ from 7-branes / singular fibers over (complex) codimension-one loci in $\mathfrak{B}$.
The factors $\mathbb{Z}_{n_i}$, associated to one of the $s$ free generators of the Mordell--Weil group, are always embedded in one of the $U(1)$ factors \cite{Cvetic:2017epq}.
We will ignore these factors, and focus on the finite factors $\mathbb{Z}_{N_t}$ generated by the torsional sections, which are embedded entirely in the non-Abelian factors $G_{\text{sc},i}$ \cite{Aspinwall:1998xj,Mayrhofer:2014opa}.

The divisors $\hat{S}^{(N)} \in H_{d-2}(Y_d)$ associated to an $N$-torsional section $S^{(N)}$ satisfy
\begin{align}
N \big( \hat{S}^{(N)} - \hat{S}_0 - \pi^{-1} (\delta) \big) = \sum_{a} \lambda_a \sigma_a \equiv \sum_{i} \sum_{b=1}^{\text{rank}(\mathfrak{g}_i)} \lambda_{i,b} \sigma_{i,b} \,, \quad \text{with} \enspace \lambda_{i,b} \in \mathbb{Z} \,,
\label{eq:Ftors}
\end{align}
where we have re-grouped the resolution divisors $\sigma_a$ on the right-hand side into their corresponding simple non-Abelian algebra $\mathfrak{g}_i$.
The term $\pi^{-1} (\delta)$ denotes a vertical divisor (i.e., pull-back of a base divisor $\delta$) that depends on the intersection properties between $\hat{S}^{(N)}$ and $\hat{S}_0$, which will not affect the discussion below.
This shows that the elements $\big(\hat{S}^{(N)} - \hat{S}_0 - \pi^{-1} (\delta) \big)$ are torsional up to the contribution of the resolution divisors $\sigma_{i,b}$, which in general dimensions are $\bbP^1$-fibered, with fiber class $\gamma_{i,b}$, over a divisor in $\mathfrak{B}$.
The coefficients $\lambda_{i,b}$ are determined by the so-called Shioda map \cite{10.3792/pjaa.65.268,2001math.....12259W,Park:2011ji,Morrison:2012ei} as follows.
The section $\hat{S}^{(N)}$ intersects at most \emph{one} of the rational fibers $\gamma_{i,b}$ of the divisors $\sigma_{i,b}$, say, $\gamma_{i,k}$, once, i.e., $\langle \hat{S}^{(N)} , \gamma_{i,b} \rangle = \delta_{k,b}$.
Then, we have
\begin{align}\label{eq:coefficients_shioda}
  \frac{\lambda_{i,b}}{N} = \sum_{c = 1}^{\text{rank}(\mathfrak{g}_i)} \langle \hat{S}^{(N)} , \gamma_{i,c} \rangle \, (C_{(i)})^{-1}_{cb} = (C_{(i)})^{-1}_{kb} \, , \quad \text{where } (C_{(i)})_{bc} = -\langle \sigma_{i,b} , \gamma_{i,c} \rangle
\end{align}

It is the existence of the element $\big(\hat{S}^{(N)} - \hat{S}_0 - \pi^{-1} (\delta) \big) \in H_{d-2} (Y_d)$ that restricts the global realization of the gauge group and accordingly the allowed spectrum of charged dynamical fields. 
By M-/F-theory duality, matter states in F-theory descend to M2-branes wrapping 2-cycles, which must have integer intersection pairing with elements in $H_{d-2}(Y_d)$, the existence of the divisor
\begin{align}\label{eq:shioda_map_relation}
\big(\hat{S}^{(N)} - \hat{S}_0 - \pi^{-1} (\delta) \big) = \tfrac{1}{N} \sum_a \lambda_a \sigma_a \,,
\end{align}
imposes, due to the fractional pre-factors $\tfrac{\lambda_a}{N}$, non-trivial constraints on the intersection numbers of 2-cycles with the divisors $\sigma_a$, which in turn determine the $\mathfrak{g}$-representation in which the matter transforms in.
Hence, \eqref{eq:shioda_map_relation} can be interpreted as an element in the cocharacter lattice, which enforces the non-trivial global structure $\pi_1(G) \cong \text{cocharacters} / \text{coroots}$ \cite{Mayrhofer:2014opa}.

While the above results are derived in compact models, the relationship between the monodromy reduction and the invariant torsion points on the generic fiber exist also in local models $Y$. 
If every singular fiber induces a monodromy in a congruence subgroup of $SL(2,\bbZ)$, then, as explained above, $Y$ has some torsional sections.
In a local geometry, sections are non-compact divisors, i.e., sit in $H_{d-2}(Y, \partial Y)$.
A relationship of the form \eqref{eq:shioda_map_relation} then implies that $\hat{S}^{(N)} - \hat{S}_0$ represents a torsional element in $H_{d-2}(Y, \partial Y)/\{\jmath_{d-2}(\sigma_a)\} \subset \Lambda$ in \eqref{eq:mag-charges_general}.\footnote{We have suppressed the vertical part $\pi^{-1}(\delta)$ here to reduce cluttering.
In general, this can be also decomposed into a compact $\pi^{-1}(\delta_c)$ and non-compact piece $\pi^{-1}(\delta_{nc})$.
The torsional generator for $H_{d-2}(Y, \partial Y)/\text{im}(\jmath_{d-2})$ is then $\hat{S}^{(N)} - \hat{S}_0 - \pi^{-1}(\delta_{nc})$.
}

This can be most easily seen for $d=4$, i.e., F-theory compactifications to eight dimensions.
Consider, for concreteness, a single $I_N$ fiber, corresponding to $\mathfrak{g} = \mathfrak{su}(N)$.
In this case, $\sigma_a = \gamma_a$, and the generic fiber $\mathfrak{f}$, which satisfies $\langle \mathfrak{f} , \mathfrak{f} \rangle = \langle \mathfrak{f} ,\sigma_a \rangle = 0$, form a basis of $H_2(Y_4)$.
The monodromy around the fiber preserves $N$-torsional points, which in the (resolved) $I_N$ fiber are situated on one of the $N$ fiber components ($\sigma_a$ and the affine component, $\sigma_0 := \mathfrak{f} - \sum_{a=1}^{N-1} \sigma_a$) each.
By ``fibering'' each point over the non-compact base $\mathfrak{B}$, we obtain a non-compact 2-cycle $\hat{S}^{(N)}_k$.
That is, they each define a class in $H_2(Y_4, \partial Y_4) \cong \text{Hom}(H_2(Y_4), \bbZ)$, characterized by the ``intersection'' with $\sigma_a$, $\hat{S}^{(N)}_k : \sigma_a \mapsto \delta_{a,k}$ for $0 \leq a,k \leq n-1$, which also implies $\hat{S}_k^{(N)}(\mathfrak{f}) = 1$ for any $k$.
Note that the zero-section is the one meeting the affine node, i.e., $\hat{S}_0 = \hat{S}^{(N)}_0$. 
With $\langle \sigma_a, \sigma_b \rangle = -C_{ab}$, it is straightforward to check that, for $k \neq 0$,
\begin{align}
\begin{split}
  & \sum_{b=1}^{N-1} (C^{-1})_{kb} \, \langle \sigma_b, \sigma_a \rangle = \delta_{k,a} = \left(\hat{S}_k^{(N)} - \hat{S}_0 \right)(\sigma_a)\, , \qquad a=1,...,N-1 \, , \\
  & \sum_{b=1}^{N-1} (C^{-1})_{kb} \, \langle \sigma_b, \mathfrak{f} \rangle = 0 = \left(\hat{S}_k^{(N)} - \hat{S}_0 \right)(\mathfrak{f}) \, ,
\end{split}
\end{align}
showing that
\begin{align}\label{eq:shioda_as_relative_cycle}
  \hat{S}_k^{(N)} - \hat{S}_0= \sum_{b=1}^{N-1} \underbrace{(C^{-1})_{kb}}_{\equiv \lambda_b / n} \, \langle \sigma_b , \cdot \rangle \in \text{Hom}(H_2(Y_4),\bbZ) \cong H_2(Y_4, \partial Y_4) \, .
\end{align}
The coefficients $\lambda_b$ are precisely as defined in \eqref{eq:coefficients_shioda};
since $(-C)^{-1}$ is the inverse Cartan matrix of $SU(N)$, $N$-times any of its entries is integral, thus showing \eqref{eq:Ftors}.\footnote{For non-compact elliptic surfaces, the vertical part $\pi^{-1}(\delta) \cong m \times \mathfrak{f}$ for some $m \in \bbZ$ always defines a trivial map, $\langle \mathfrak{f}, \cdot \rangle = 0 \in \text{Hom}(H_2(Y_4), \bbZ)$.}
Since $\langle \sigma_c, \cdot \rangle = \jmath_{d-2}(\sigma_c)$ in \eqref{eq:j_map_general}, relations of the sort \eqref{eq:Ftors} directly identify the torsional section (more precisely, the linear combination $\hat{S}^{(N)}_k - \hat{S}_0$) as a representative higher-form symmetry charges \eqref{eq:boundary_torsion_rep_I_N}.
Note that this expression also agrees with the relationship between the asymptotic and root junctions \eqref{eq:k_frac_junc}, serving as further proof that the two concepts are equivalent.

\subsubsection*{Torsional Sections in the Boundary Homology}

A relationship of the form \eqref{eq:shioda_as_relative_cycle} implies that the 2-cycle $\hat{S}^{(N)}_k - \hat{S}_0$ maps to a torsion element in $H_2(Y_4, \partial Y_4)/\text{im}(\jmath_2) \cong \text{im}(\partial_2) \subset H_1(\partial Y)$.
To see this explicitly, consider the points, $z_\text{tors}$ and $z_0$, marked by the two sections $S^{(N)}_k$ and $S_0$, respectively, on a reference $T^2$ fiber $\mathfrak{f}_p$ of the boundary fibration $\partial Y \rightarrow \partial \mathfrak{B} \cong S^1$.
Then, $z_\text{tors}$ traces out the 1-cycle $\partial_2(\hat{S}^{(N)}_k) = \hat{S}^{(N)}_k|_{\partial Y}$ on $\partial Y$, as we move it through the family of fibers over the base $S^1$; the same applies to $z_0$ tracing out $\partial_2(\hat{S}_0) = \hat{S}_0|_{\partial Y}$.
In any individual fiber, the two points are homologous.
But, by encircling the base $S^1$ once, the monodromy $K$ ``twists'' the section $S^{(N)}_k$ around the zero-section.
This twist corresponds to the 1-cycle $\big( \hat{S}^{(N)}_k - \hat{S}_0 \big)\big|_{\partial Y} = {\cal C} \in H_1(\partial Y)$.

To quantify this twist, we can use the standard presentation of the torus as $\bbC / (m \, \tau_p + n )$, with $z_0 \mapsto 0$, where $\tau_p$ is the complex structure of the torus $\mathfrak{f}_p$.
Then, every point on the torus can be represented as $\left( \begin{smallmatrix} x \\ y \end{smallmatrix} \right) \equiv x \, \tau_p+ y + (m \, \tau_p + n)$ with $0 \leq x,y < 1$.
The monodromy map $K$ acts via matrix multiplication, $\left( \begin{smallmatrix} x \\ y \end{smallmatrix} \right) \mapsto K \left( \begin{smallmatrix} x \\ y \end{smallmatrix} \right)$, which fixes $z_0 = \left( \begin{smallmatrix} 0 \\ 0 \end{smallmatrix} \right)$.
In the covering space $\bbC$ of the torus, this defines a translation by $(K-{\bf 1}) \left( \begin{smallmatrix} x \\ y \end{smallmatrix} \right)$, which on the quotient $\bbC / (m \,  \tau_p+ n)$ corresponds to a 1-chain ${\cal C}$.

Being preserved under the monodromy now precisely means that $z_\text{tors} \equiv \left( \begin{smallmatrix} x_t \\ y_t \end{smallmatrix} \right)$ maps onto itself in $\bbC / (m \, \tau_p+ n)$.
That is, the chain ${\cal C} = a {\cal A} + b {\cal B}$ is a 1-\emph{cycle} on $\mathfrak{f}_p$, expressed in terms of the ${\cal A}$ and ${\cal B}$ cycles, with coefficients given by $(K - {\bf 1}) \left( \begin{smallmatrix} x_t \\ y_t \end{smallmatrix} \right) = \left( \begin{smallmatrix} a \\ b \end{smallmatrix} \right)$ with $a,b \in \bbZ$.
Finally, the fact that $z_\text{tors}$ is an $N$-torsional point means that $(x_t, y_t) = (\chi/N, \upsilon/N)$ for some $\chi, \upsilon \in \{0,...,N-1\}$ (see, e.g., \cite{diamond2006first}).\footnote{By definition, the Mordell--Weil group law for sections in elliptic fibration is just the fiberwise addition of points, the latter of which can be represented as the ``usual'' addition in $\bbC / (m\tau_p + n)$. This makes the given presentation of the torsion points apparent.}
Therefore, we see from $N \left( \begin{smallmatrix} a \\ b \end{smallmatrix} \right) = (K - {\bf 1}) \left( \begin{smallmatrix} N x_t \\ N y_t \end{smallmatrix} \right) = (K - {\bf 1}) \left( \begin{smallmatrix} \chi \\ \upsilon \end{smallmatrix} \right)$ that $N {\cal C} \in \text{im}(K - {\bf 1}) = \text{im}(\kappa)$.
From \eqref{eq:boundhom}, we see that ${\cal C} = \big( \hat{S}^{(N)}_k - \hat{S}_0 \big)\big|_{\partial Y}$ indeed represents an $N$-torsional element in $\text{coker}(\kappa) \subset H_1(\partial Y)$.

\section{Anomalies of 1-Form Center Symmetries in M-theory}
\label{sec:anomalies}

The defect group structure represents an 't Hooft anomaly between the electric 1-form and magnetic $(D - 3)$-form symmetry \cite{Freed:2006ya,Freed:2006yc,Albertini:2020mdx}.
Another such potential anomaly involving the 1-form center symmetry arises in spacetime dimension $D \geq 5$ \cite{Apruzzi:2020zot,Cvetic:2020kuw,BenettiGenolini:2020doj}, as a generalization of the ``anomaly in the coupling-space'' \cite{Cordova:2019jnf,Cordova:2019uob} in $D=4$, where a non-trivial background field for the 1-form center symmetry affects the periodicity of the theta angle \cite{Aharony:2013hda}.
In $D\geq 5$ dimensions, this turns into a genuine mixed 't Hooft anomaly between 1-form center symmetries and $(D-5)$-form $U(1)_I$ instanton symmetries.\footnote{
The anomaly restricts possible gaugings of center symmetries --- i.e., it affects physically allowed global gauge groups --- whenever $U(1)_I$ must be gauged for consistency \cite{Apruzzi:2020zot,Cvetic:2020kuw}.}

In this section, we discuss the origin of the anomaly in gauge theories from M-theory compactifications on Calabi--Yau spaces $Y_d$ ($\dim_\mathbb{R} Y_d \equiv d = 4,6$).
As we will see, one way to derive the anomaly is to reduce the 11d Chern--Simons term in the presence of boundary fluxes that parametrize the 1-form symmetry background.
Similar to the discussion in Section \ref{sec:3app}, the computation is performed in the Abelian phase, i.e., on the Coulomb branch of the ${\cal N}=1$ gauge theory in 7d or 5d, which corresponds to a desingularized internal space $Y_d$.
We then interpret the result in the singular / non-Abelian limit, as well as in cases that admit an 8d / 6d F-theory description.
Note that there could be counterterms / topological sectors which (partly) cancel this anomaly field-theoretically.
These will not be captured by our analysis of the Chern--Simons term in the presence of boundary fluxes, but could manifest in other aspects of the M-theory geometry, see \cite{Apruzzi:2021vcu} for a recent discussion.

\subsection{Background Fields for 1-Form Symmetries in M-theory}

Consider M-theory on a spacetime ${\cal M} = M_{11-d} \times Y_d$, where the $d$-dimensional ``internal'' space $Y_d$ is non-compact with asymptotic boundary $\partial Y_d$.
Assuming that $M_{11-d}$ has a topologically trivial boundary, boundary fluxes of the M-theory 3-form potential $C_3$ are then encoded in fluxes on $\partial Y_d$.
More precisely, dual to \eqref{eq:long_seq_rel_hom}, there is a long exact sequence,
\begin{align}\label{eq:long_seq_cohomology}
\ldots \rightarrow H^n({\cal M}, \partial {\cal M}) \stackrel{\hat\jmath_n^*}{\rightarrow} H^n({\cal M}) \stackrel{\hat\imath^*_n}{\rightarrow} H^n(\partial {\cal M}) \rightarrow H^{n+1}({\cal M}, \partial {\cal M}) \rightarrow \ldots
\end{align}
involving the relative cohomology $H^n({\cal M}, \partial {\cal M})$.
A non-trivial boundary flux of $C_3$ corresponds to an element in $\text{im}(\hat\imath^*_4) \subset H^4(\partial {\cal M})$, where $\hat\imath^*_n$ is the map on $n$-forms induced by the natural inclusion $\hat\imath: \partial {\cal M} \rightarrow {\cal M}$, and is the cohomological version of $\imath_n$ in \eqref{eq:long_seq_rel_hom}.
With ${\cal M} = M_{11-d} \times Y_d$ and the assumption that $M_{11-d}$ is closed, i.e., $\partial {\cal M} = M_{11-d} \times \partial Y_d$, the boundary fluxes are encoded in the map
\begin{align}
\begin{aligned}
	H^4({\cal M}) \cong \bigoplus_{p+q=4} H^p(M_{11-d}) & \otimes H^q(Y_d) &&\stackrel{\imath^*_4}{\longrightarrow} && \bigoplus_{p+q=4} H^p(M_{11-d}) \otimes H^q(\partial Y_d) \cong H^4(\partial {\cal M})\, , \\
	F & \otimes \omega && \mapsto && F \otimes \imath_q^*(\omega) \, ,
\end{aligned}
\end{align}
with $\imath^*_q: H^q(Y_d) \rightarrow H^q(\partial Y_d)$ the analogous map in the long exact sequence \eqref{eq:long_seq_cohomology} associated to the relative cohomology for $\partial Y_d \subset Y_d$.

In the following, we focus on $p=q=2$, as gauge fields $A_a$ in $M_{11-d}$ arise from the Kaluza--Klein decomposition of the M-theory 3-form potential,
\begin{align}\label{eq:C3_KK_decomp}
	C_3 = C^{(M)}_3 + \sum_{\mathrm{w}_a \in H^2(Y_d)} A_a \wedge \mathrm{w}_a \, ,
\end{align}
where $C^{(M)}_3$ is a 3-form in $M_{11-d}$.
If we only include $\mathrm{w}_a = \jmath_2^*(\omega_a) \in H^2_{\text{cpt}}(Y_d) \equiv \text{im}(\jmath_2^*) = \text{ker}(\imath^*_2)$ with compact support, the associated $A_a$'s correspond to the Cartan $U(1)$s of dynamical gauge symmetries.
The flux, or the field strength, of such a configuration is then
\begin{align}
	G_4 = G_4^{(M)} + \sum_{\omega_a \in H^2(Y_d, \partial Y_d)} F_a \otimes \jmath_2^*(\omega_a) \, ,
\end{align}
where $F_a$ represents the first Chern-class of a line bundle in $H^2(M_{11-d})$, and corresponds to the field strength of the gauge field $A_a$.
In the following, we will assume $G_4^{(M)}=0$, as we are only interested in contributions to 2-form backgrounds in $M_{11-d}$.

Non-trivial ``boundary'' fluxes are labelled by elements in $ H^2(Y_d) / \text{im}(\jmath_2^*) = H^2(Y_d) / \text{ker}(\imath^*_2)$ $\cong \text{im}(\imath^*_2) \subset H^2(\partial Y_d)$ \cite{Morrison:2020ool}.
They can be represented by classes $\tilde\omega_k \in H^2(Y_d)$ with $\imath^*_2(\tilde\omega_k) \neq 0$.
``Turning on'' these boundary fluxes means that we include additional terms
\begin{align}\label{eq:G_4_with_boundary_fluxes}
	G_4 = \sum_{a} F_a \otimes \jmath_2^*(\omega_a) + \sum_k B_k \otimes \tilde\omega_k\, .
\end{align}
These additional terms can be related to the electrically and magnetically charged objects, \eqref{eq:1-form-charges_general} and \eqref{eq:mag-charges_general}, of the higher-form symmetry.
By virtue of the commutative diagram via Poincar\'e--Lefschetz duality (see, e.g., \cite{hatcher2002algebraic}),
\begin{equation}\label{eq:comm_diag}
	\begin{tikzcd}[row sep = normal, column sep= normal  ]
		\ldots \arrow[r] & H^1(\partial Y_d) \arrow{r}{}{d_1} \arrow{d}{}{\cong} & H^2(Y_d, \partial Y_d) \arrow{r}{}{\jmath_2^*} \arrow{d}{}{\cong} & H^2(Y_d) \arrow{r}{}{\imath_2^*} \arrow{d}{}{\cong} & H^2(\partial Y_d) \arrow{d}{}{\cong} \arrow[r] & \ldots \\
		\ldots \arrow[r] & H_{d-2}(\partial Y_d) \arrow{r}{}{\imath_{d-2}} & H_{d-2}(Y_d) \arrow{r}{}{\jmath_{d-2}} & H_{d-2}(Y_d, \partial Y_d) \arrow{r}{}{\partial_{d-k}} & H_{d-3}(\partial Y_d) \arrow[r] & \ldots
	\end{tikzcd}
\end{equation}
we have $H^2(Y_d) / \text{im}(\jmath^*_2) \cong H_{d-2}(Y_d, \partial Y_d) / \text{im}(\jmath_{d-2}) = \Lambda_\text{mag.} \cong \bbZ^f \oplus \Gamma$ from \eqref{eq:explicit_formula_Gamma}.

Thus, the additional terms $B_k \otimes \tilde{\omega_k}$ in $G_4$ arrange into
\begin{align}\label{eq:free_and_torsion_fluxes}
	 H^2(M_{11-d}) \otimes (\mathbb{Z}^f \oplus \Gamma) \cong H^2(M_{11-d})^{\otimes f} \oplus H^2(M_{11-d}; \Gamma) \, .
\end{align}
Contributions in $H^2(M_{11-d})^{\otimes f}$ correspond to background gauge fields of flavor symmetries in $M_{11-d}$.
As they are in the free part of $H^2(\partial Y_d) \cong H_{d-3}(\partial Y_d)$, they have trivial linking pairing\footnote{This is dual to the pairing \eqref{eq:linking_pairing_homology} in the boundary homology.} with any other boundary flux, and hence commutes with any other flux background.
We will return to these backgrounds later, and first focus on the torsional part $H^2(M_{11-d}; \Gamma)$, which physically correspond to background fields for global 1-form $\Gamma$ symmetries in $M_{11-d}$.

Turning on a background flux in $H^2(M_{11-d}; \Gamma) \ni b \equiv B \otimes \tilde\omega$ corresponds to a \emph{torsional} internal flux $\tilde\omega_t \in \Gamma \subset H^2(Y_d) / \text{ker}(\imath^*_2)$, i.e., $\imath_2^*(\tilde\omega_t) \neq 0 \in H^2(\partial Y_d)$, but $N \, \tilde\omega_t \in \text{ker}(\imath_2^*) = \text{im}(\jmath_2^*)$ for some $N \in \mathbb{N}_0$.
This means that there is an integer linear combination
\begin{align}\label{eq:linear_relation_boundary_flux}
	N \, \tilde\omega_t = \sum_a (S^{-1})_{ta} \, \jmath_2^*(\omega_a) \equiv \sum_a \lambda_a \, \jmath_2^*(\omega_a) \, ,
\end{align}
which is Poincar\'e-dual to the homology relation \eqref{eq:lin_rel_homology}.
Since $\tilde\omega_t$ is only defined modulo $\text{ker}(\imath^*_2)= \text{im}(\jmath_2^*)$, we can restrict $\lambda_a \in \{0,...,N-1\}$.
Thus, we can formally write
\begin{align}\label{eq:G4_shifted_U1s}
	G_4 = \sum_a F_a \otimes \jmath_2^*(\omega_a) + B \otimes \tilde\omega_t = \sum_a \left( F_a + \frac{\lambda_a}{N} B \right) \otimes \jmath_2^*(\omega_a) \, ,
\end{align}
which can be interpreted as a $N$-fractional shift of the Cartan fluxes by the 1-form symmetry.
This interpretation agrees with the field theoretic description of 1-form symmetry transformation in the Abelian phase of the gauge theory \cite{Hsin:2018vcg,Cordova:2019uob}.
Note that only the $N$-fractional part of the shift to the Cartan fluxes in \eqref{eq:G4_shifted_U1s} is well-defined, since the boundary flux $\tilde\omega_t$ is only defined modulo $H^2_\text{cpt}(Y_d)$.

The presentation \eqref{eq:G4_shifted_U1s} has the advantage that we can ``straightforwardly'' perform the usual KK-reduction of the M-theory Chern--Simons term 
\begin{align}\label{eq:11d-CS-term}
	\frac16 \int_{\cal M} C_3 \wedge G_4 \wedge G_4 \, .
\end{align}
By that, we mean that the integral is strictly speaking only defined for compactly supported cohomology forms on non-compact spaces.
Presumably, a mathematically more rigorous definition of this coupling in terms of differential cohomology classes \cite{Witten:1996hc,Witten:1999vg,Belov:2006jd,Belov:2006xj,Fiorenza:2012mr,Fiorenza:2012ec}, which we will not attempt to utilize here, can encompass contributions from both compactly supported and boundary fluxes.
For the $N$-torsional fluxes that parametrize the 1-form symmetry backgrounds, \eqref{eq:G4_shifted_U1s} allows us to circumvent this process and evaluate the integrals of products of the compactly supported 2-forms $\jmath_2^{*} (\omega_a)$, albeit with the fractional coefficients.
As we will see, this approach is sufficient to derive the \emph{fractionalization}, i.e., a fractional shift of the instanton density of gauge theories in the presence of a 1-form symmetry background that matches field theory results.

\subsection{Compactification to 7d}
\label{subsec:7d_anomalies}

Let us apply the above results to $\dim Y_d \equiv d = 4$, i.e., M-theory compactified to seven dimensions.
In the case the ansatz \eqref{eq:C3_KK_decomp} for $C_3$ includes only compactly supported fluxes in $Y_4$, the reduction of the 11d Chern--Simons term \eqref{eq:11d-CS-term} produces the term
\begin{align}\label{eq:CS-term-7d}
	& \frac{1}{6} \int_{{\cal M}_{11}} C_3 \wedge G_4 \wedge G_4 = \frac12 \sum_{a,b} \int_{M_{7}} C_3^{(M)} \wedge  F_a \wedge F_b \times \int_{Y_4} \jmath_2^*(\omega_a) \wedge \jmath_2^*(\omega_b) \, ,
\end{align}
where the factor of 3 comes from the assumption that the boundary of the 7d spacetime $M_7$ is trivial, allowing for Stoke's theorem on $G_4^{(M)} = dC_3^{(M)}$.
If $F_a$ are the Cartan $\mathfrak{u}(1)$s of a non-Abelian gauge symmetry $\mathfrak{g}$ in $M_7$, then $\int_{Y_4}\jmath_2^*(\omega_a) \wedge \jmath_2^*(\omega_b) \equiv -C_{ab}$ is the (negative) Cartan matrix of $G$.
Moreover, it coincides with the matrix $M_{ai}$ in \eqref{eq:smith_decomp}, where the basis $\sigma_a \in H_{d-2}(Y_4) = H_2(Y_4)$ and $\gamma_i \in H_2(Y_4)$ is formed by the 2-cycles dual to $\omega_a \in H^2(Y_4)$ in both cases.
In the non-Abelian limit (i.e., when we blow-down the compact curves dual to $\omega_a$ in $Y_4$), this term produces the 7d-coupling
\begin{align}\label{eq:7d_instanton_coupling}
	\sum_{a,b} \int_{M_{7}} \frac{(-C_{ab})}{2} \, C_3^{(M)} \wedge  F_a \wedge F_b \rightarrow \int_{M_7} C_3^{(M)} \wedge \text{Tr}(F^2)
\end{align}
between the instanton density $\text{Tr}(F^2)$\footnote{The trace is normalized such that a 1-instanton configuration integrates to an integer over any integer 4-cycle in $M_7$.} of $G$ and the 3-form $C_3^{(M)}$.

Including an $N$-torsional boundary flux $\tilde\omega_t$, and its fractional shift \eqref{eq:G4_shifted_U1s} it induces on the Cartan $\mathfrak{u}(1)$s, the coupling becomes\footnote{We have implicitly used the ``continuum description'' \cite{Kapustin:2014gua,Gaiotto:2014kfa,Gaiotto:2017yup} for the 1-form background gauge field $B$ as an ordinary differential form, for which the wedge product makes sense.
Regarding $B \in H_2(M_D; \Gamma)$ as a differential cohomology class, one should replace $B \wedge B$ by the Pontryagin square operation $\mathfrak{P}(B)$.
}
\begin{align}\label{eq:CS-term_reduced_to_7d}
\begin{split}
	& \sum_{a,b} \frac{(-C_{ab})}{2} \int_{M_7} C_3^{(M)} \wedge \left( F_a + \frac{\lambda_a}{N} B \right) \wedge \left( F_b + \frac{\lambda_b}{N} B \right) \\
	= & \sum_{a,b} \int_{M_7} C_{ab} \, C_3^{(M)} \wedge \left(\frac12 F_a \wedge F_b + \frac{\lambda_a}{N} F_b \wedge B + \frac{\lambda_a \lambda_b}{2N^2} B \wedge B \right) .
\end{split}
\end{align}
With $-C_{ab} \equiv M_{ab}$ in \eqref{eq:smith_decomp}, we see from \eqref{eq:lambda_via_smith_decomp} that 
\begin{align}
	\sum_a C_{ab} \frac{\lambda_a}{N} = \sum_{a,c,j} S_{ac} \, D_{cj} \, T_{jb} \frac{(S^{-1})_{ta}}{N} = \sum_j \frac{D_{tj} \, T_{jb}}{N} = \frac{n_t}{N} T_{tb} = T_{tb} \in \bbZ \, ,
\end{align}
because $N \equiv n_t$ is the torsion order of the boundary flux $\tilde\omega_t$ that we turned on.
This means that the cross terms $C_3^{(M)} \wedge F_b \wedge B$ in \eqref{eq:CS-term_reduced_to_7d} actually have integer coefficients.

Since $F_a$ and $B$ are all integer 2-forms (more precisely, 2-cocycles) in $M_7$, we see that a non-trivial 1-form symmetry background corresponding to $\tilde\omega_t$ leads to a shift
\begin{align}
	\sum_{a,b} \frac{-C_{ab}}{2} \, F_a \wedge F_b + \sum_{a,b} (-C_{ab}) \frac{\lambda_a \, \lambda_b}{2N^2} B \wedge B + \text{integer contributions} \, .
\end{align}
In the non-Abelian limit, the instanton coupling \eqref{eq:7d_instanton_coupling} thus becomes
\begin{align}\label{eq:shift_instanton_general}
	\int_{M_7} C_3^{(M)} \wedge \text{Tr}(F^2) \rightarrow \int_{M_7} C_3^{(M)} \wedge \Big( \text{Tr}(F^2) + \frac{1}{2N} \underbrace{\sum_{a,b} (-C_{ab}) \frac{\lambda_a \lambda_b}{N}}_{= -\sum_b T_{tb} (S^{-1})_{tb} \in \bbZ} B \wedge B + \text{integer 4-form} \Big) .
\end{align}

This fractional shift leads to an 't Hooft anomaly between the 1-form center symmetry, and the large gauge transformations of the $U(1)$ symmetry $C_3^{(M)} \rightarrow C_3^{(M)} + \Lambda^{(M)}_3$, where $\Lambda^{(M)}_3$ is a closed 3-form \cite{Apruzzi:2020zot,Cvetic:2020kuw,BenettiGenolini:2020doj}.
While $C_3^{(M)}$ is a background field in 7d when gravity is decoupled, in supergravity, it becomes the dynamical field dual of the anti-symmetric 2-tensor in the gravity multiplet, and as such must enjoy an unbroken $U(1)$ symmetry.
A mixed anomaly with a 1-form symmetry thus prevents the gauging of this 1-form symmetry, and thus restricts possible $\pi_1(G)$, if the 1-form symmetry corresponds to a center symmetry.
It is straightforward to uplift this to term and the anomaly to 8d, where $C_3^{(M)}$ now becomes a 4-form gauge potential $B_4$ coupling to the instanton density, with analogous implications for global gauge group structures in 8d supergravity \cite{Cvetic:2020kuw}.

\paragraph{Example} Consider M-theory on $Y_4 = \mathbb{C}^2/\bbZ_N$, which gives rise to a 7d theory with $G = SU(N)$.
The corresponding exact sequence in relative homology \eqref{eq:comm_diag} collapses in this case to a short exact sequence (see, e.g., \cite{Garcia-Etxebarria:2019cnb}),
\begin{align}\label{eq:relative_hom_sequence_C2/Gamma}
\begin{split}
	& 0 \longrightarrow H_2(Y) \stackrel{\jmath_2}{\longrightarrow} H_2(Y, \partial Y) \stackrel{\partial_2}{\longrightarrow} H_1(\partial Y) \rightarrow 0 \, , \\
	\text{with} \quad &  H_2(Y, \partial Y) \cong \text{Hom}(H_2(Y) , \bbZ) \, , \quad H_1(\partial Y) \cong \bbZ_N \, .
\end{split}
\end{align}
For $Y_4 = \mathbb{C}^2/\bbZ_N$, it is well-known that $H_2(Y)$ is spanned by $N-1$ $\bbP^1$'s ($\bbP^1_a$, $a=1,...,N-1$), which intersect each other in the form of an $SU(N)$ Dynkin diagram, that is, 
\begin{align}\label{eq:suN_cartan_matrix_7d}
	C_{ab} = \langle \bbP^1_a , \bbP^1_b \rangle = 
	\begin{pmatrix}
		-2 & 1 & 0 & \ldots & 0 \\
		1 & -2 & 1 & \ddots & 0 \\
		0 & 1 & -2 & \ddots & \vdots \\
		\vdots & \ddots & \ddots & \ddots \\
		0 & 0 & \ldots & 1 & -2
	\end{pmatrix} \, .
\end{align}
A Smith decomposition $C = S D T$ yields
\begin{align}\label{eq:smith_decomp_suN_cartan}
\begin{split}
	& S = 
	\begin{pmatrix}
		1 & \ldots & 1 & \ldots & \ldots & 1 \\
		\vdots & 2 & 2 & \ldots & \ldots & 2 \\
		1 & 2 & 3 & \ldots & \ldots & 3 \\
		\vdots & \vdots & 3 & \ddots & & \vdots \\
		\vdots & \vdots & \vdots & & N-2 & N-2\\
		1 & 2 & 3 & \ldots & N-2 & N-1
	\end{pmatrix}^{-1} \, , \quad 
	T = 
	\begin{pmatrix}
		1 & 0 & \ldots & 0 & 1 \\
		0 & 1 & \ddots & \vdots & 2  \\
		\vdots & \ddots & \ddots &  & \vdots \\
		0 & 0 & \ldots & 1 & N-2 \\
		0 & 0 & \ldots & & 1
	\end{pmatrix} \, ,\\
	& D = \text{diag}[\underbrace{-1, -1, \ldots,  -1, -N}_{N-1}] \, ,
\end{split}
\end{align}
confirming $\text{Tor}(H_2(Y, \partial Y)/\text{im}(\jmath_2)) \cong \bbZ_N = H_1(\partial Y)$ in \eqref{eq:relative_hom_sequence_C2/Gamma}, corresponding to the $(N-1)$-th entry in the diagonal matrix $D$.
Hence, the coefficients in \eqref{eq:shift_instanton_general} are
\begin{align}
	\lambda_a = (S^{-1})_{N-1, a} = a \, .
\end{align}
As a cross check, we find the standard identity for $SU(N)$ Cartan matrices,
\begin{align}\label{eq:sum_cartan_a}
	\sum_{a=1}^{N-1} (-C_{ab}) \frac{\lambda_a}{N} = \sum_{a=1}^{N-1} (-C_{ab}) \frac{a}{N}= \begin{cases}
		0 \in \bbZ \, , \quad b = 1,..., N-2 \, ,\\
		1 \in \bbZ \, , \quad b = N - 1 \, .
	\end{cases}
\end{align}

Thus, a background flux $B$ for the $\bbZ_N$ 1-form symmetry induces shift \eqref{eq:G4_shifted_U1s} of the Cartan fluxes given by
\begin{align}
	F_a \rightarrow F_a + \frac{a}{N} B
\end{align}
which agrees with the action of the 1-form symmetry in the maximally Abelian phase of the gauge theory \cite{Hsin:2018vcg,Cordova:2019uob}.
Moreover, it also leads to the fractional shift
\begin{align}\label{eq:instanton_shift_suN_7d}
	\frac{1}{2N} \sum_{a,b=1}^{N-1} (-C_{ab}) \frac{\lambda_a \lambda_b}{N} B \wedge B = \frac{N-1}{2N} B \wedge B \, ,
\end{align}
to the instanton density \eqref{eq:shift_instanton_general}.
This agrees with the field theoretic results about the fractionality of $SU(N)$ instantons in the presence of a 1-form center background field \cite{Kapustin:2014gua,Gaiotto:2014kfa,Gaiotto:2017yup,Hsin:2018vcg,Cordova:2019uob}.

\paragraph{Models with F-theory uplift}
The result \eqref{eq:shift_instanton_general} applies, mutatis mutandis, to M-theory on elliptically fibered $Y_4$.
These models can be interpreted as an $S^1$-reduction of F-theory compactified on $Y_4$.
If this 8d theory has gauge symmetry $\mathfrak{g}$, then, in 7d, there are $\text{rank}(\mathfrak{g}) + 1$ compact 2-cycles $\sigma_a \in H_2(Y_4)$.
The additional 2-cycle is the generic fiber $\mathfrak{f}$ of $\pi: Y_4 \rightarrow B_2$, and gives rise to the vector multiplet in 7d obtained by integrating the Ramond-Ramond 2-form field in 8d\footnote{In a type IIB description, this 2-form field is the reduction of the 10d RR-field $C_4$ on the base $B_2$.} over the $S^1$.
Because $\mathfrak{f}$ has intersection number 0 with any compact 2-cycle in $Y_4$, it does not contribute to the Chern--Simons term \eqref{eq:CS-term-7d}.
In the presence of a boundary flux, the shifted 7d Chern--Simons terms \eqref{eq:shift_instanton_general} thus are equivalent to a fractional shift of the $G$-instantons inherited from an 8d 1-form background field.

As a concrete example, consider $Y_4$ the neighborhood of an $I_N$ fiber, which realizes an $\mathfrak{su}(N)$ gauge symmetry in 8d F-theory.
The set of compact curves, $\{\sigma_a\}$, $a=1,...,N$, intersect in the affine $SU(N)$ Dynkin diagram:
\begingroup\makeatletter\def\f@size{10}\check@mathfonts
\begin{align}
\begin{split}
	&\langle \sigma_a , \sigma_b \rangle
	= \left(\begin{array}{ccccc|c}
		& & & & & 1 \\
		& & & & & 0 \\
		& & C & & & \vdots\\
		& & & & & 0 \\
		& & & & & 1 \\ \hline
		1 & 0 & \hdots & 0 & 1 & -2
	\end{array}
	\right) = \left( \begin{array}{cc|c}
		& &  0 \\
		\multicolumn{2}{c|}{S^{-1}}  & \vdots \\
		& & 0 \\ \hline
		-1 & \hdots & -1
	\end{array} \right)
	\left( \begin{array}{c|c}
		D & 0\\ \hline
		0 & 0
	\end{array} \right)
	\left( \begin{array}{ccc|c}
		& & & -2 \\
		& & & -3 \\
		& T & & \vdots \\
		& & &  1-N \\ 
		& & & -1 \\ \hline
		0 & \hdots & 0 & 1
	\end{array} \right),
\end{split}
\end{align}
\endgroup
with the $(N-1) \times (N-1)$ matrices $(S, D, T)$ given in \eqref{eq:smith_decomp_suN_cartan}.
As expected, $\jmath_2((S^{-1})_{N,b} \,  \sigma_b) = -\sum_{a=1}^N \sigma_a = -\mathfrak{f}$ has trivial intersection with any compact 2-cycle $\sigma_c$.
The remaining $N-1$ 2-cycles gives rise to the same structure as the $\mathfrak{su}(N)$ example from $Y_4 = \mathbb{C}^2 / \bbZ_N$ above, with the $N$-torsional boundary flux given by $\tilde\sigma_t = \tfrac{1}{N} (S^{-1})_{N-1,c} \, \sigma_c = \sum_{c=1}^{N-1} \tfrac{c}{N} \sigma_c$.
Analogously, the fractional shift of the instanton density of the $SU(N)$ symmetry is \eqref{eq:instanton_shift_suN_7d}, which is the same as the shift in 8d \cite{Cvetic:2020kuw}.

\subsection{Compactification to 5d}

In compactifications on $Y_6$ to five dimensions, a reduction analogous to \eqref{eq:CS-term_reduced_to_7d} of the M-theory Chern--Simons term with the ansatz \eqref{eq:C3_KK_decomp} give rise to the 5d Chern--Simons terms \cite{Cadavid:1995bk,Ferrara:1996hh,Ferrara:1996wv}
\begin{align}\label{eq:CS-term_reduced_to_5d}
	\frac16 \sum_{\alpha,\beta,\gamma} \int_{M_5} K_{\alpha \beta \gamma} \, A_\alpha \wedge F_\beta \wedge F_\gamma \, .
\end{align}
When we only consider dynamical gauge fields in 5d spacetime (for which we use Latin indices $(\alpha,\beta,\gamma) \rightarrow (a,b,c)$), the internal pieces $\mathrm{w}_{\alpha,\beta,\gamma}$ in \eqref{eq:C3_KK_decomp} are all compactly supported 2-forms $\jmath_2^*(\omega_a)$.
In this case, the coefficients $K_{abc} = \int_{Y_6} \jmath_2^*(\omega_a) \wedge \jmath_2^*(\omega_b) \wedge \jmath_2^*(\omega_c)$ have a natural interpretation as the integral of products of compactly supported 2-forms, or, dually, as intersection number of 4-cycles $\sigma_a \in H_4(Y_6)$.
Physically, $K_{abc}$ encode the Coulomb branch dynamics of the effective 5d gauge theory \cite{Morrison:1996xf,Intriligator:1997pq}.

In the following, we are interested in the terms with $\omega_\alpha \equiv \tilde\omega_I \in H^2(Y_6)$ fixed, such that $\imath^*_2(\tilde\omega_I) \in \mathbb{Z}^f \subset H^2(\partial Y_6)$ (cf.~formula \eqref{eq:free_and_torsion_fluxes}), and let the indices $(\beta,\gamma) \rightarrow (b,c)$ run over compactly supported 2-forms that span the Cartan subgroup of the gauge group $G$.
As stated above, $\tilde{F}_I \otimes \tilde\omega_I$ corresponds to a background Cartan $U(1)_I$ flux of the global 0-form symmetries.
Since the Poincar\'e--Lefschetz-dual 4-cycle is a relative homology class, $\text{PD}(\tilde\omega_I) = \epsilon_I \in H_4(Y_6, \partial Y_6)$, that is \emph{not} in the image of $\jmath_4$, it may be regarded as a non-compact 4-cycle in $Y_6$.
In general, the corresponding Chern--Simons coefficients $K_{Ibc}$ encode mass parameters of the $G$ gauge theory \cite{Jefferson:2018irk}.
Crucially, the global 0-form symmetry include $U(1)$ factors that charge the instanton particles of the $G$ gauge theory, which can enhance the flavor symmetry (the part of the global symmetry charging hypermultiplets of the effective gauge theory) at the UV fixed point \cite{Seiberg:1996bd}.
For our discussion, we focus on the effective gauge theory phase, in which $\tilde\omega_I$ corresponds to an instantonic $U(1)_I$ global symmetry, rather than a Cartan $U(1)$ of the (classical) flavor symmetry.

Passing, for convenience, to the Poincar\'e--Lefschetz-dual homology description, we have $\text{PD}(\omega_{b,c}) = \sigma_{b,c} \in H_4(Y_6)$, and we can form the intersection product $\sigma_b \cdot \sigma_c \equiv \gamma_{bc} \in H_2(Y_6)$, which yields a 2-cycle in $Y_6$.
On the other hand, $\tilde\omega_I \in H^2(Y_6)/\text{im}(\jmath_2^*)$ is represented by an element in $H^2(Y_6)$, which we abusively also denote by $\tilde\omega_I$.
Then $\text{PD}(\tilde\omega_I) = \epsilon_I \in H_4(Y_6, \partial Y_6) \cong \text{Hom}(H_2(Y_6), \bbZ)$.
This now gives a straightforward way to ``define'' $K_{Ibc} = \epsilon_I (\gamma_{bc})$.

In the 5d Chern--Simons terms \eqref{eq:CS-term_reduced_to_5d}, turning on the 1-form symmetry background \eqref{eq:G4_shifted_U1s} leads to
\begin{align}\label{eq:fractional_shift_CS-terms_5d}
\begin{split}
	& \frac16 \sum_{\alpha,\beta,\gamma} \int_{M_5} K_{\alpha \beta \gamma} \, A_\alpha \wedge F_\beta \wedge F_\gamma \supset \frac12 \sum_{b,c} \int_{M_5} K_{Ibc} \, A_I \wedge F_b \wedge F_c \\
	\longrightarrow \quad & \frac12 \sum_{b,c} \int_{M_5}  K_{Ibc} \, A_I \wedge \left( F_b \wedge F_c + \frac{2\lambda_b}{N} F_c \wedge B + \frac{\lambda_b \lambda_c}{N^2} B \wedge B \right) \, .
\end{split}
\end{align}
To match this with the field theory results, one needs to show that the cross terms $F_c \wedge B$ again have integer coefficients.
Arguing for the integrality analogously to the 7d case requires the intersection pairing between the basis of $(d-2)$- and 2-cycles.
However, the spaces $H_4(Y_6)$ and $H_2(Y_6)$ are in general very different, and the intersection product $H_4(Y_6) \times H_4(Y_6) \rightarrow H_2(Y_6)$ depends on details of $Y_6$.
As such, it is difficult to make a general argument that applies to all geometries.
Instead, we will look at concrete examples, where we can calculate \eqref{eq:fractional_shift_CS-terms_5d} explicitly.

\subsubsection{Examples with 5d UV Fixed Point}

Let us illustrate this first in a simple example, namely, for the rank one 5d ${\cal N}=1$ theory with gauge symmetry $G = SU(2)$, theta angle $\theta = 0$, and no matter.
The latter two conditions ensure that the 1-form $\mathbb{Z}_2$ center symmetry is not explicitly broken by any matter or instanton particles \cite{Morrison:2020ool,Albertini:2020mdx}.
This gauge theory has as (continuous) global 0-form $U(1)$ instanton symmetry, which enhances to an $SU(2)$ at the UV fixed point \cite{Seiberg:1996bd,Morrison:1996xf,Intriligator:1997pq}.
The (non-compact) Calabi--Yau threefold $Y_6$ that describes this theory via M-theory is local neighborhood of an $\mathbb{F}_0 \cong \bbP^1 \times \bbP^1 \equiv \sigma$ surface, which generates $H_4(Y_6) \cong \bbZ$.
Furthermore, $H_2(Y_6) \cong \bbZ^2$ is generated by two $\bbP^1$s, $\gamma_1$ and $\gamma_2$, inside $\sigma$, with intersection pairing $\langle \sigma, \gamma_1 \rangle = \langle \sigma, \gamma_2 \rangle = -2$.
The corresponding Smith decomposition \eqref{eq:smith_decomp} is simple,
\begin{align}
	(-2, -2) = (1) (-2, 0) \begin{pmatrix} 1 & 1 \\ 0 & 1 \end{pmatrix} \, ,
\end{align}
implying that $\epsilon_1 = \eta_1 + \eta_2 = \tfrac12 \jmath_2^*(\sigma)$ is the generator of the $\Gamma \cong \mathbb{Z}_2$ 1-form symmetry backgrounds (and so $\lambda_b \equiv \lambda_\sigma = 1$ and $N=2$ in \eqref{eq:fractional_shift_CS-terms_5d}).
Meanwhile, the ``non-compact divisor'' $\epsilon_I \in H_4(Y_6, \partial Y_6) \cong \text{Hom}(H_2(Y_6),\bbZ)$ corresponding to the $U(1)_I$ global symmetry is given by
\begin{align}
	\epsilon_I = \eta_2: H_2(Y_6) \rightarrow \bbZ \, , \quad a_1\gamma_1 + a_2\gamma_2 \mapsto a_2 \, .
\end{align}
Additionally, we need $\sigma \cdot \sigma = -2(\gamma_1 + \gamma_2)$.
This means that $K_{Ibc} \equiv K_{I, \sigma, \sigma} = \epsilon_I(\sigma \cdot \sigma) = -2$.

In the presence of the $\bbZ_2$ 1-form symmetry background, the shifted Chern--Simons term \eqref{eq:fractional_shift_CS-terms_5d} then becomes
\begin{align}
\begin{split}
	& \frac12 \int_{M_5} \sum_{b,c} K_{Ibc} \, A_I \wedge \left( F_b \wedge F_c + \frac{2\lambda_b}{N} F_c \wedge B + \frac{\lambda_b \lambda_c}{N^2} B \wedge B \right) \\
	= & - \int_{M_5} A_I \wedge \left( F_\sigma \wedge F_\sigma + F_\sigma \wedge B + \frac14 B \wedge B \right) \\
	\stackrel{\text{non-ab.~limit}}{\longrightarrow} \quad & - \int_{M_5} A_I \wedge \left( \text{Tr}(F_{SU(2)}^2) + \frac{1}{4} B \wedge B + \text{integer terms} \right)\, .
\end{split}
\end{align}
The mixed 't Hooft anomaly between $U(1)_I$ and the center 1-form symmetry of $SU(2)$ resulting from this fractional shift of the instanton density indeed agrees with expectations from field theory \cite{BenettiGenolini:2020doj}.

Note that this geometric computation is strictly speaking only valid on the Coulomb branch of the 5d theory.
While the extrapolation to the sublocus, where we have an effective non-Abelian gauge theory, agrees with previous work, our approach cannot preclude a cancellation of this anomaly through a topological sector which is hidden on the Coulomb branch.
Indeed, recent work \cite{Apruzzi:2021vcu} suggests the existence of such sectors on the Higgs branch, which geometricly can only be accessed by passing through the strongly-coupled SCFT point via deformation.
Because of this, we do not make any claims about how anomaly lifts to the UV theory.

In general, there is no reason to expect that modifications from the effective field theory expectations can only arise from the Higgs branch.
Indeed, certain UV-effects can also be found on the Coulomb branch, which indicates a more subtle effect of turning on 1-form symmetry backgrounds in the SCFT.
For that, we consider setups realizing pure $SU(N \geq 3)_k$ gauge theories with Chern--Simons level $2-N < k < N-2$.
These theories have an $\bbZ_{\text{gcd}(N,k)}$ 1-form symmetry, and rank $f=1$ global symmetry given by the instantonic $U(1)_I$ \cite{Morrison:2020ool,Albertini:2020mdx}.
A possible M-theory geometry is a local Calabi--Yau neighborhood $Y_6$ of $N-1$ Hirzebruch surfaces $\sigma_a \cong \bbF_{n_a}$ that intersect transversely in a chain.
Leaving the detailed computation to appendix \ref{app:smith_decomp_suN_5d}, we present here the shift of the Chern--Simons terms by the $\bbZ_{\text{gcd}(N,k)}$-torsional boundary flux:
\begin{align}\label{eq:instanton_shift_5d_suN}
\begin{split}
	& \frac12 \sum_{a,b} \int_{M_5}  K_{Iab} \, A_I \wedge \left( F_a \wedge F_b  + \frac{\lambda_a \lambda_b}{\text{gcd}(N,k)^2} B \wedge B \right) \\
	= & \int_{M_5} \left( \sum_{a,b}  \frac{Q_{22}}{2} (-C_{ab}) A_I \wedge F_a \wedge F_b - \frac{Q_{21}(N+k)}{2} A_I \wedge F_{N-1} \wedge F_{N-1} \right) \\
	+ & \int_{M_5} \frac{N-1}{2N} \ell^2 \left( Q_{22}  - Q_{21} (N+k) \frac{N-1}{N} \right) A_I \wedge B \wedge B \, ,
\end{split}
\end{align}
where $\ell = N/\text{gcd}(N,k) \mod N$ is the generator of the $\bbZ_{\text{gcd}(N,k)}\subset \bbZ_{N}$ subgroup of the center of $SU(N)$.
The coefficients $Q_{22}$ and $Q_{21}$ are integers fixed by a Euclidean Algorithm on $(N, k-N)$:
\begin{align}\label{eq:relation_Qs}
  \text{gcd}(N, k-N) = \text{gcd}(N,k) = Q_{22} N - Q_{21} (k-N)\, .
\end{align}

This result does not immediately agree with the expectations from the effective $SU(N)_k$ gauge description.
Neglecting the contribution proportional to $Q_{21} A_I \wedge F_{N-1} \wedge F_{N-1}$ above, one would interpret the first term, $\sum_{a,b}  \frac{Q_{22}}{2} (-C_{ab}) A_I \wedge F_a \wedge F_b$, as the Coulomb branch expression of $A_I \wedge (Q_{22} \text{Tr}(F^2))$.
That is, instanton density of $SU(N)$ is coupled to $U(1)_I$ with charge $Q_{22}$.
Then --- again neglecting the term proportional to $Q_{21}$ --- the last line in \eqref{eq:fractional_shift_CS-terms_5d} would precisely correspond to the fractional shift of $SU(N)/\bbZ_{\text{gcd}(N,k)}$ instantons.
Thus, the $Q_{21}$-terms are expected to be non-perturbative corrections to the effective $SU(N)_k$ gauge description.
Furthermore, we have not fully explored the invariance of \eqref{eq:relation_Qs} under $\big( Q_{12}, Q_{22} \big) \rightarrow \big(Q_{21} + m \tfrac{N}{\text{gcd}(N,k)} , Q_{22} + m \tfrac{k-N}{\text{gcd}(N,k)} \big)$, though this seems to be related to a redefinition of the generator for $U(1)_I$, cf.~\eqref{eq:def_top_U1_suN_5d_example}.
To gain a better understanding of these terms, it would be instructive to find a field theoretic description of such non-perturbative corrections, and/or verify the geometric result from another construction of the 5d SCFT that is the UV-completion of the $SU(N)_k$ gauge theory.

\subsubsection{5d KK-Theories and 6d Anomalies}

If $Y_6$ is elliptically fibered (over a K\"ahler manifold $\mathfrak{B}_4$), then M-theory on $Y_6$ gives rise to a so-called 5d KK theory.
The UV-completion of such a gauge theory is not a genuine 5d SCFT, but rather a 6d SCFT on an $S^1$ (hence the name).
In this reduction, the 5d 1-form symmetry receives contributions from both 1-form and 2-form symmetries in 6d \cite{Morrison:2020ool}.
Moreover, the instanton density of the 6d gauge symmetry couples via a Green--Schwarz mechanism to dynamical tensor fields, which on an $S^1$ reduce to vector multiplets associated to additional $U(1)$ (0-form) gauge symmetries in 5d.
Therefore, in the 5d KK-theory, the 6d mixed anomaly between the 1-form symmetry and the large gauge transformations of tensor fields \cite{Apruzzi:2020zot} are encoded in the Chern--Simons terms $K_{abc}$ involving three compact divisors.

For an example, consider the 6d non-Higgsable $SU(3)$ theory.
The corresponding 5d KK-theory is M-theory compactified on an elliptically fibered $\pi: Y_6 \rightarrow \mathfrak{B}_4$, given by the Calabi--Yau neighborhood of three intersecting $\bbF_1$ surfaces \cite{DelZotto:2017pti}:\footnote{
The basis of 2-cycles on each $\bbF_1$ surface is $\{f,e\}$, which on $\bbF_1$ intersect as $e \cdot e = -1, e \cdot f =1, f \cdot f = 0$.
}
\begin{equation}
	\begin{tikzcd}[row sep = normal, column sep= normal  ]
		\sigma_1 \cong \bbF_1 \arrow[dash]{r}{e} \arrow[dash]{d}{e} & \bbF_1 \cong \sigma_2 \\
		\sigma_3 \cong \bbF_1 \arrow[dash]{ur}{e} &
	\end{tikzcd} \, .
\end{equation}
The intersection $\sigma_1 \cdot \sigma_2 = \sigma_2 \cdot \sigma_3 = \sigma_1 \cdot \sigma_3 = e$ is the $(-1)$-curve in each $\sigma_a$, which is the section of the $\bbP^1$-fibration on $\sigma_a$ with fiber $f_a$.
The generic elliptic fiber is $\mathfrak{f} = f_1 + f_2 + f_3$ in homology.
Thus, all $\sigma_a$ are fibered over the same genus-0 curve $C \subset B_4$, which has self-intersection number $-3$ inside $\mathfrak{B}_4$.
Furthermore, $\sigma_a \cdot \sigma_a = -2(e + 3f_a)$.

A basis of $H_2(Y_6)$ is given by $\gamma_i \in \{f_1, f_2, f_3, e\}$.
From the intersection matrix,
\begingroup\makeatletter\def\f@size{10}\check@mathfonts
\begin{align}\label{eq:smith_decomp_su3_NHC}
	M_{ai} = \langle \sigma_a , \gamma_j \rangle = \begin{pmatrix}
		-2 & 1 & 1 & -1 \\
		1 & -2 & 1 & -1 \\
		1 & 1 & -2 & -1
	\end{pmatrix}
	=
	\underbrace{\begin{pmatrix}
		1 & 1 & 0\\
		1 & 2 & 0\\
		1 & 1 & 1
	\end{pmatrix}^{-1}}_{S^{-1}}
	\begin{pmatrix}
		- 1 & 0 & 0 & 0\\
		0 & -3 & 0 & 0\\
		0 & 0 & -3 & 0
	\end{pmatrix}
	\underbrace{\begin{pmatrix}
		1 & 1 & -2 & 2 \\
		0 & 1 & -1 & 1 \\
		0 & 0 & 0 & 1 \\
		0 & 0 & 1 & -1
	\end{pmatrix}}_{T},
\end{align}
\endgroup
we see that the surfaces $\sigma_{1,2}$ form the Cartan $U(1)$s of the $SU(3)$ gauge symmetry, whose $\bbZ_3$ 1-form center symmetry is encoded in the boundary flux (cf.~\eqref{eq:lin_rel_homology})
\begin{align}\label{eq:boundary_flux_su3_NHC}
	\tilde\sigma_t = \frac13 \sum_a \lambda_a \jmath_4(\sigma_a) \equiv \frac{1}{3} \sum_a (S^{-1})_{2,a} \jmath_4(\sigma_a) = \frac{1}{3} (\jmath_4(\sigma_1) + 2 \jmath_4(\sigma_2)) \, .
\end{align}
The additional dynamical $U(1)_\text{d}$ gauge field is dual to $\xi_3 = (S^{-1})_{3,a} \sigma_a = \sigma_1 + \sigma_2 + \sigma_3 = \pi^{-1}(C)$, which is a vertical divisor in $Y_6$, and hence corresponds to the $S^1$-reduction of the 6d tensor field.

Thus, the coefficients of the relevant Chern--Simons terms,
\begin{align}\label{eq:CS-term-SU3-NHC}
	\frac12 \sum_{a,b =1}^2 \int_{M_5} K_{\text{d}ab} \, A_\text{d} \wedge F^{(SU(3))}_a \wedge F^{(SU(3))}_b \, ,
\end{align}
are $K_{\text{d} ab} = \langle \xi_3 , \sigma_a \cdot \sigma_b \rangle = 3 (-C^{(SU(3))})_{ab}$, with $(-C^{(SU(3))})$ the Cartan matrix of $SU(3)$.
In the limit where we collapse the fibers $f_{1,2}$ in $Y_6$, and thereby enhancing the gauge symmetry to $SU(3) \times U(1)_\text{d}$, these Chern--Simons terms become
\begin{align}
	-3 \int_{M_5} A_\text{d} \wedge \text{Tr}(F^{(SU(3))} \wedge F^{(SU(3))}) \, ,
\end{align}
indicating that $U(1)_\text{d}$ gauges the instanton symmetry of $SU(3)$ ``with charge 3'' \cite{Morrison:2020ool}.
This also agrees with the reduction of the corresponding 6d Green--Schwarz coupling with tensor charge $3$.
Furthermore, this charge prefactor also ensures that the 1-form center symmetry has no mixed anomaly with the large gauge transformations of $U(1)_\text{d}$, since the fractional shift induced by the boundary flux \eqref{eq:boundary_flux_su3_NHC} is
\begin{align}
	\sum_{a,b=1}^2 K_{\text{d} ab} \, \frac{\lambda_a \lambda_b}{2 \cdot 3^2} \, B \wedge B = \frac{1}{6} \sum_{a,b=1}^{2}  a \, b \,(-C^{(SU(3))})_{ab} \, B \wedge B = B \wedge B \, .
\end{align}
This matches the absence of the corresponding 1-form anomaly in 6d \cite{Apruzzi:2020zot}.

Let us further consider $\tilde\rho := \eta_4 \in \text{Hom}(H_2(Y_6), \bbZ) \cong H_4(Y_6, \partial Y_6)$, which from \eqref{eq:smith_decomp} satisfies $-\jmath_4(\xi_3) = 3 \eta_4$, i.e., is the generator of the second $\bbZ_3$ factor in the 5d 1-form symmetry.
Since $\tilde\rho(f_a)=0$, $\tilde\rho(e)=1$, $\tilde\rho$ can be interpreted as a non-compact \emph{vertical} divisor in $Y_6$, whose projection $\pi(\tilde\rho)$ onto $\mathfrak{B}_4$ intersects the compact $(-3)$-curve $C$ once.
In the 6d F-theory setting, wrapping D3-branes on $\pi(\tilde\rho)$ gives rise to string-like surface defects that are charged under the 6d $\bbZ_3$ 2-form symmetry \cite{DelZotto:2015isa,Morrison:2020ool}.

In 5d, we can now study the effects of turning on background fields $B$ for the $\bbZ_3$ 1-form center symmetry of $SU(3)$, as well as $B_\text{d}$ for the $\bbZ_3$ 1-form symmetry that descends from the 6d 2-form symmetry.
While the first corresponds to the shift $F_a \rightarrow F_a + \frac{a}{3} B$ for the $SU(3)$ Cartan fluxes, the second shifts the field strength of the $U(1)_\text{d}$ as $F_\text{d} \rightarrow F_\text{d} + \frac13 B_\text{d}$.
The latter can be viewed as a transformation $A_\text{d} \rightarrow A_\text{d} + \frac13 \epsilon$, where $\epsilon$ is a flat $\bbZ_3$ connection \cite{Cordova:2019jnf,Cordova:2019uob,BenettiGenolini:2020doj}.
From \eqref{eq:CS-term-SU3-NHC}, we would then obtain the term
\begin{align}
	\frac12 \sum_{a,b=1}^2 K_{\text{d}ab} \, \frac13 \epsilon \wedge \frac{a}{3} B \wedge \frac{b}{3} B = \frac13 \epsilon \wedge B \wedge B \, .
\end{align}
This constitutes a mixed 't Hooft anomaly between the two $\bbZ_3$ 1-form symmetries, which can be written in terms of a 6d anomaly theory,
\begin{align}
  {\cal A}[B_\text{d}, B] = \int_{X_6} \frac13 B_\text{d} \wedge B^2 \, ,
\end{align}
where $\partial X_6 = M_5$ is an auxiliary manifold whose boundary is the 5d spacetime.

It appears natural to uplift this anomaly to the 6d gauge theory that corresponds to F-theory compactified on $Y_6$.
Here, this would be a mixed 't Hooft anomaly between the 6d 2-form $\bbZ_3$ symmetry for the instanton strings, and the 1-form $\bbZ_3$ center symmetry of the non-Higgsable $SU(3)$ gauge sector.
One intuitive explanation for this anomaly is that, by turning on a background field for the 1-form center symmetry, the instanton number fractionalizes, being now instantons of an $SU(3)/\bbZ_3$ bundle.
Compared to the instanton strings of $SU(3)$, which have charge $0 \mod 3$ under the $\bbZ_3$ 2-form symmetry, the $SU(3)/\bbZ_3$ instantons have charge $1 \mod 3$, and hence screen all asymptotic charges of the 2-form symmetry.
It would be interesting to investigate potential field theoretic counterterms for this anomaly, and, if present, their imprints in geometry.

\section*{Acknowledgments}

We thank Fabio Apruzzi and Miguel Montero for valuable discussions.
M.C.~and H.Y.Z.~are supported in part by DOE Award DE-SC013528Y.
M.C.~further acknowledges support by the Simons Foundation Collaboration grant \#724069 on ``Special Holonomy in Geometry, Analysis and Physics'', the Slovenian Research Agency (ARRS No.~P1-0306), and the Fay R.~and Eugene L.~Langberg Endowed Chair.

\begin{appendix}

\section{Defect Group Structures of M-theory on Elliptic Fibrations}
\label{app:defect_group_elliptic}

In this appendix, we compute the defect group structure for the 7d theories listed in Table \ref{tab:fibers_and_kappa}.
The strategy is to first determine a representative of the $N_i$-torsional 1-cycles ${\cal C}_i$ in $H_1(\partial Y)_\text{tors} \cong \text{coker}(\kappa)$ as a linear combination of the ${\cal A} = (1,0)$ and ${\cal B} = (0,1)$ cycles, which intersect on $T^2$ as ${\cal A}^2 = {\cal B}^2 = 0$ and ${\cal A} \cdot {\cal B} = - {\cal B} \cdot {\cal A} = 1$.
This can be done by a Smith decomposition on $\kappa$.
Then, there is a 2-chain $\Sigma_i \subset \partial Y$ with $\partial \Sigma_i = N_i {\cal C}_i$.
The linking pairing $L: \text{Tor}(H_1(\partial Y)) \times \text{Tor}(H_1(\partial Y)) \rightarrow \mathbb{Q}/\bbZ$ is then computed as
\begin{align}
  L({\cal C}_i, {\cal C}_j) = \frac{1}{N_i} \Sigma_i \cdot {\cal C}_j \mod \bbZ \, .
\end{align}

\paragraph{$I_N$ fiber:}
The standard representation of the monodromy around an $I_N$ fiber is given by $K = \left( \begin{smallmatrix} 1 & -N \\ 0 & 1 \end{smallmatrix} \right)$.
It is easy to see that
\begin{align}\label{eq:gamma_I_N_fiber}
	\kappa = \begin{pmatrix} 1 & 0 \\ 0 & 1 \end{pmatrix} - \begin{pmatrix} 1 &-N \\ 0 & 1 \end{pmatrix} = \begin{pmatrix} 0 & N \\ 0 & 0 \end{pmatrix}
\end{align}
has $\text{im}(\kappa) \cong \bbZ$, generated by the 1-cycle $\kappa {\cal B} = \left( \begin{smallmatrix} N \\ 0 \end{smallmatrix} \right) = N \times {\cal A}$.
Therefore,
\begin{align}
	\text{coker}(\kappa) = \frac{H^1(T^2) }{\text{im}(\kappa)} \cong \frac{\langle \left(\begin{smallmatrix} 1 \\ 0 \end{smallmatrix}\right) \rangle_\bbZ \times \langle \left(\begin{smallmatrix} 0 \\ 1 \end{smallmatrix}\right) \rangle_\bbZ }{\langle \left(\begin{smallmatrix} N \\ 0 \end{smallmatrix}\right) \rangle_\bbZ} \cong \bbZ_N \times \bbZ \, ,
\end{align}
where $\langle v \rangle_\bbZ$ denotes the $\bbZ$-span of a vector $v$.
Here, the $\bbZ_N$ factor is generated by the ${\cal A}$-cycle, and the free $\bbZ$ factor by the ${\cal B}$-cycle.
Another way to read \eqref{eq:gamma_I_N_fiber} is that the 2-cycle $\Sigma \subset \partial Y$, obtained by tracing the ${\cal B}$-cycle over any point $p \in S^1_\text{base}$ once around the circle $S^1_\text{base}$ back to $p$, has boundary $\partial \Sigma = K{\cal B} - {\cal B} = -\kappa {\cal B} = -N \, {\cal A}$.
This allows us to compute the linking pairing:
\begin{align}\label{eq:defect_pairing_I_N}
	L({\cal A}, {\cal A}) = -\tfrac{1}{N} \Sigma \cdot {\cal A} \mod \bbZ = \tfrac{1}{N} \mod \bbZ \, ,
\end{align}
where we have used the specific realization of $\Sigma$ as the ${\cal B}$-cycle fibered over $S^1$, which intersects the ${\cal A}$-cycle only in one fiber, where ${\cal B} \cdot_{T^2} {\cal A} = -1$.
This result agrees with the physical expectation that the $I_N$ fiber realizes an $\mathfrak{su}(N)$ gauge symmetry as a 7d M-theory compactification, whose electric 1-form and magnetic 4-form symmetry are both $\mathbb{Z}_N$, with the defect group pairing \eqref{eq:defect_pairing_I_N}.

\paragraph{$I^*_{N-4}$ fiber:} The monodromy around an $I_{N-4}^*$, $N\geq 4$, type fiber, in the above $(\cal A, B)$ basis, is $K = \left( \begin{smallmatrix} -1 & N-4 \\ 0 & -1 \end{smallmatrix} \right)$.
To compute $\text{coker}(\kappa)$, it is easiest to distinguish between even and odd $N$.

For even $N \equiv 2n$, $n \geq 2$, we have
\begin{align}\label{eq:smith_decomp_D2n_monodromy}
	\kappa =\begin{pmatrix} 1& 0 \\ 0 & 1 \end{pmatrix} - \begin{pmatrix} -1 & 2n - 4 \\ 0 & -1 \end{pmatrix} = \begin{pmatrix} 2 & 4-2n\\ 0 & 2 \end{pmatrix} = \begin{pmatrix} 1& 2-n \\ 0 & 1 \end{pmatrix} \begin{pmatrix} 2 & 0 \\ 0 & 2 \end{pmatrix} \begin{pmatrix} 1 & 0 \\ 0 & 1 \end{pmatrix} \, ,
\end{align}
where the last equality can be interpreted as a Smith decomposition, showing in particular that
\begin{align}
	\text{coker}(\kappa) \cong \frac{\langle \left(\begin{smallmatrix} 1 \\ 0 \end{smallmatrix}\right) \rangle_\bbZ \times \langle \left(\begin{smallmatrix} 0 \\ 1 \end{smallmatrix}\right) \rangle_\bbZ}{\langle \left( \begin{smallmatrix} 2 \\ 0 \end{smallmatrix} \right) \rangle_\bbZ \times \langle \left( \begin{smallmatrix} 0 \\ 2 \end{smallmatrix} \right) \rangle_\bbZ } \cong \bbZ_2 \times \bbZ_2 \, .
\end{align}
Now consider the 2-chain $\Sigma_1 \subset \partial Y$ obtained from fibering the 1-cycle $-{\cal A} \equiv \left( \begin{smallmatrix} -1 \\ 0 \end{smallmatrix} \right)$ once over the boundary circle back to a reference point $p \in S^1_\text{base}$.
Then we have $\partial \Sigma_1 = -\kappa (-{\cal A}) = 2 {\cal A}$, showing that ${\cal A}$ is a 2-torsional 1-cycle in $\partial Y$.
Likewise, we can also see that ${\cal B}$ is a 2-torsional cycle in $\partial Y$, as $-\kappa \left( \begin{smallmatrix} 2-n \\ -1 \end{smallmatrix} \right) = \left( \begin{smallmatrix} 0 \\ 2 \end{smallmatrix} \right)$, i.e., $2{\cal B}$ is the boundary of the 2-chain $\Sigma_2$ swept out by moving the 1-cycle $\left( \begin{smallmatrix} 2-n \\ -1 \end{smallmatrix} \right)$ around $S^1_\text{base}$ once.
To compute the linking pairing, first note that 
\begin{align}
	\begin{aligned}
		& \Sigma_1 \cdot {\cal A} = -{\cal A} \cdot_{T^2} {\cal A} = 0 \, , \quad && \Sigma_1 \cdot {\cal B} = -{\cal A} \cdot_{T^2} {\cal B} = -1 \, , \\
		& \Sigma_2 \cdot {\cal A} = ((2-n) {\cal A} - {\cal B}) \cdot_{T^2} {\cal A} = -1\, , \quad && {\cal B} \cdot \Sigma_2 = {\cal B} \cdot_{T^2} ((2-n) {\cal A} - {\cal B} = 2-n \, .
	\end{aligned}
\end{align}
Thus, the linking pairing, in the basis $({\cal A}, {\cal B})$, is (with $\tfrac12 = -\tfrac12 \mod \bbZ$)
\begin{align}\label{eq:linking_pairing_Deven}
	\text{even $N$} : \quad L_{I^*_{N-4}} = \begin{pmatrix} 0 & \tfrac12 \\ \tfrac12 & \tfrac{N}{4} \end{pmatrix} \mod \bbZ \, .
\end{align}

For odd $N = 2n+1$, $n\geq 2$, we have
\begin{align}\label{eq:smith_decomp_Dodd_monodromy}
	\gamma =\begin{pmatrix} 1& 0 \\ 0 & 1 \end{pmatrix} - \begin{pmatrix} -1 & 2n - 3 \\ 0 & -1 \end{pmatrix} = \begin{pmatrix} 2 & 3-2n\\ 0 & 2 \end{pmatrix} = \begin{pmatrix} 1& n-1 \\ -2 & 3-2n \end{pmatrix} \begin{pmatrix} 1 & 0 \\ 0 & 4 \end{pmatrix} \begin{pmatrix} 0 & -1 \\ 1 & 2 \end{pmatrix} \, ,
\end{align}
implying $\text{coker}(\kappa) \cong \bbZ_4$.
A generating 4-torsional 1-cycle is given by $\left( \begin{smallmatrix} 1-n \\ 1 \end{smallmatrix} \right) = \tfrac14 \times (- \kappa \left( \begin{smallmatrix} 1 \\ -2 \end{smallmatrix} \right))$, i.e., four copies of it is the boundary of the 2-chain $\Sigma$ obtained from moving $\left( \begin{smallmatrix} 1 \\ -2 \end{smallmatrix} \right)$ around $S^1_\text{base}$ once.
From $\Sigma \cdot \left( \begin{smallmatrix} 1-n \\ 1 \end{smallmatrix} \right) = \left( \begin{smallmatrix} 1 \\ -2 \end{smallmatrix} \right) \cdot_{T^2} \left( \begin{smallmatrix} 1-n \\ 1 \end{smallmatrix} \right) = 3 -2n = 4 -N$, we find the linking pairing
\begin{align}\label{eq:linking_pairing_Dodd}
	\text{odd $N$}: \quad L_{I^*_{N-4}} = \tfrac{-N}{4} \mod \bbZ \, .
\end{align}

Again, this reproduces the physical expectations, namely that $I^*_{N-4}$ realizes an $\mathfrak{so}(2N)$ gauge symmetry.
For even $N = 2n$, the electric and magnetic symmetries are $\bbZ_2 \times \bbZ_2$, whose defect group pairing is \eqref{eq:linking_pairing_Deven}.\footnote{
Note that this deviates from \cite{Albertini:2020mdx} for odd $n$, or $N \not\equiv 0 \mod 4$, where the pairing matrix is given by $\left( \begin{smallmatrix} 1/2 & 0\\ 0 & 1/2\end{smallmatrix} \right)$.
One can easily check that we arrive in this form by a simply basis change $({\cal A}, {\cal B}) \rightarrow ({\cal A + B}, {\cal B})$.}
Meanwhile, for odd $N=2n+1$, $\mathfrak{so}(2N)$ has $\bbZ_4$ center symmetry, which pairs with its magnetically dual $\bbZ_4$ symmetry as \eqref{eq:linking_pairing_Dodd}.

\paragraph{$IV$ fiber:} The monodromy is given by $K=\left( \begin{smallmatrix} -1 & -1\\ 1& 0\end{smallmatrix} \right)$, so
\begin{align}
	\kappa = \begin{pmatrix} 2 & 1 \\ -1 & 1 \end{pmatrix} = \begin{pmatrix} 2 & -1 \\ -1 & 1 \end{pmatrix} \begin{pmatrix} 1 & 0 \\ 0 & 3 \end{pmatrix} \begin{pmatrix} 1 & 2 \\ 0 & 1 \end{pmatrix} \, .
\end{align}
Therefore $\text{coker}(\kappa) \cong \bbZ_3$, whose generator can be represented by the ${\cal B}$-cycle, since $3{\cal B} = \left( \begin{smallmatrix} 0 \\ 3 \end{smallmatrix} \right) = -\kappa \left( \begin{smallmatrix} -1 \\ 2 \end{smallmatrix} \right)$ is the boundary of $\Sigma$ obtained from fibering $\left( \begin{smallmatrix} -1 \\ 2 \end{smallmatrix} \right)$ over $S^1_\text{base}$.
From $\Sigma \cdot {\cal B} = \left( \begin{smallmatrix} -1 \\ 2 \end{smallmatrix} \right) \cdot_{T^2} \left( \begin{smallmatrix} 0 \\ 1 \end{smallmatrix} \right) = -1$, the linking pairing is $L({\cal B}, {\cal B}) = -\tfrac13 \mod \bbZ$.
This defines the linking pairing for the $\bbZ_3$ 1-form electric / 4-form magnetic symmetry of an $\mathfrak{su}(3)$ gauge symmetry, which is different than an $\mathfrak{su}(3)$ realized via an $I_3$ fiber.

\paragraph{$IV^*$ fiber:} In this case, the monodromy is $K=\left( \begin{smallmatrix} -1 & 1\\ -1& 0\end{smallmatrix} \right)$, so
\begin{align}
  \kappa = \begin{pmatrix} 2 & -1 \\ 1 & 1 \end{pmatrix} = \begin{pmatrix} 2 & -1 \\ 1 & 0 \end{pmatrix} \begin{pmatrix} 1 & 0 \\ 0 & 3 \end{pmatrix} \begin{pmatrix} 1 & 1 \\ 0 & 1 \end{pmatrix} \, .
\end{align}
Hence coker$(\kappa) \cong \bbZ_3$, which can be generated by the ${\cal A}$-cycle, since $3{\cal A} = \left( \begin{smallmatrix} 3 \\ 0 \end{smallmatrix} \right) = -\kappa \left( \begin{smallmatrix} -1 \\ 1 \end{smallmatrix} \right) = \partial \Sigma$, which also defines $\Sigma$ to be the 2-chain obtained by fibering $\left( \begin{smallmatrix} -1 \\ 1 \end{smallmatrix} \right)$ over $S^1_{\text{base}}$.
Then $\Sigma \cdot {\cal A} = \left( \begin{smallmatrix} -1 \\ 1 \end{smallmatrix} \right) \cdot_{T^2} \left( \begin{smallmatrix} 1 \\ 0 \end{smallmatrix} \right) = -1$, so the linking pairing is $L({\cal A}, {\cal A}) = -\tfrac13 = \frac23 \mod \bbZ$
Physically, this agrees with the expectation from an 7d $\mathfrak{e}_6$ gauge symmetry from an $IV^*$ fiber, which has 1-form electric / 4-form magnetic $\bbZ_3$ symmetry, and defect group pairing $-\tfrac13 = \tfrac23 \mod \bbZ$ \cite{Albertini:2020mdx}.

\paragraph{$III$ fiber:} The monodromy of this fiber is $K=\left( \begin{smallmatrix} 0 & -1\\ 1& 0\end{smallmatrix} \right)$, so
\begin{align}
	\kappa = \begin{pmatrix} 1 & 1 \\ -1 & 1 \end{pmatrix} = \begin{pmatrix} 1 & 0 \\ -1 & 1 \end{pmatrix} \begin{pmatrix} 1 & 0 \\ 0 & 2 \end{pmatrix} \begin{pmatrix} 1 & 1 \\ 0 & 1 \end{pmatrix} \, .
\end{align}
Therefore $\text{coker}(\kappa) \cong \bbZ_2$, which is generated by the ${\cal B}$-cycle, since $2{\cal B} = \left( \begin{smallmatrix} 0 \\ 2 \end{smallmatrix} \right) = -\kappa \left( \begin{smallmatrix} 1 \\ -1 \end{smallmatrix} \right)$ is the boundary of $\Sigma$ obtained from fibering $\left( \begin{smallmatrix} 1 \\ -1 \end{smallmatrix} \right)$ over $S^1_\text{base}$.
From $\Sigma \cdot {\cal B} = \left( \begin{smallmatrix} 1 \\ -1 \end{smallmatrix} \right) \cdot_{T^2} \left( \begin{smallmatrix} 0 \\ 1 \end{smallmatrix} \right) = 1$, the linking pairing is $L({\cal B}, {\cal B}) = \tfrac12 \mod \bbZ$.
Physically, this agrees with the expectation from a 7d theory with an $\mathfrak{su}(2)$ gauge symmetry from a type $III$ fiber.

\paragraph{$III^*$ fiber:} The monodromy is $K=\left( \begin{smallmatrix} 0 & 1\\ -1& 0\end{smallmatrix} \right)$, so
\begin{align}
  \kappa = \begin{pmatrix} 1 & -1 \\ 1 & 1 \end{pmatrix} = \begin{pmatrix} 1 & -1 \\ 1 & 0 \end{pmatrix} \begin{pmatrix} 1 & 0 \\ 0 & 2 \end{pmatrix} \begin{pmatrix} 1 & 1 \\ 0 & 1 \end{pmatrix} \, .
\end{align}
The coker$(\kappa) \cong \bbZ_2$ can be represented by the ${\cal A}$-cycle, which is the boundary of $\Sigma$ obtained by fibering $\left( \begin{smallmatrix} -1 \\ 1 \end{smallmatrix} \right)$ around $S^1_\text{base}$: $-\kappa \left( \begin{smallmatrix} -1 \\ 1 \end{smallmatrix} \right) = \left( \begin{smallmatrix} 2 \\ 0 \end{smallmatrix} \right)$.
The defect group pairing is determined from $\Sigma \cdot {\cal A} = \left( \begin{smallmatrix} -1 \\ 1 \end{smallmatrix} \right) \cdot_{T^2} \left( \begin{smallmatrix} 1 \\0 \end{smallmatrix} \right) = -1$ to be $L({\cal A}, {\cal A}) = -\frac12 = \frac12 \mod \bbZ$, which agrees with that of the 1-form electric / 4-form magnetic symmetry of a 7d $\mathfrak{e}_7$ theory \cite{Albertini:2020mdx}.

\paragraph{$II$ and $II^*$ fibers:} Lastly, these fiber have monodromy $K_{II} = \left( \begin{smallmatrix} 1 & -1\\ 1& 0\end{smallmatrix} \right)$ and $K_{II^*} = \left( \begin{smallmatrix} 1 & 1\\ -1& 0\end{smallmatrix} \right)$.
Since, in these cases,
\begin{align}
	\kappa_{II} = \begin{pmatrix} 0 & 1 \\ -1 & 1 \end{pmatrix} \, , \quad \kappa_{II^*} = \begin{pmatrix} 0 & -1 \\ 1 & 1 \end{pmatrix}
\end{align}
is unimodular, the cokernels are trivial, conforming with expectations of neither a trivial nor an $\mathfrak{e}_8$ gauge symmetry have any higher-form electric and magnetic symmetries.

\section{Instanton Fractionalization in 5d \texorpdfstring{\boldmath{$SU(N)_k$}}{SU(N)[k]} Theories}\label{app:smith_decomp_suN_5d}

In this appendix, compute the 't Hooft anomaly between the instanton $U(1)_I$ 0-form symmetry and the $\bbZ_{\text{gcd}(N,k)}$ 1-form symmetry of a 5d $SU(N)_k$ gauge symmetry.
The Calabi--Yau threefold geometry is the local neighborhood of the surface configuration \cite{Bhardwaj:2019ngx}
\begin{align}\label{eq:surface_config_suN}
\begin{split}
	\bbF_{N-k-2} \quad _{e_1} \overset{\line(2,0){10}}{\;}_{h_2} & \quad \bbF_{N-k-4} \quad _{e_2}\overset{\line(2,0){5}}{\;} \ldots \overset{\line(2,0){5}}{\;}_{h_m} \quad \bbF_{N-k-2m} \quad _{e_m} \overset{\line(2,0){10}}{\;}_{e_{m+1}} \quad \bbF_{-N+k+2(m+1)} \quad _{h_{m+1}} \overset{\line(2,0){5}}{\;} \ldots \\
	\ldots \overset{\line(2,0){5}}{\;}_{e_{N-2}} & \quad \bbF_{N+k-4} \quad _{h_{N-2}} \overset{\line(2,0){10}}{\;}_{e_{N-1}} \quad \bbF_{N+k-2} \, ,
\end{split}
\end{align}
where the degrees $n_a = N-k-2a$ for $a \leq m$, and $n_a = -(N-k-2a)$ for $a>m$, are all positive.
Each of these (complex) surfaces have a basis of 2-cycles, $(e_a, f_a)$, with $h_a := e_a + n_a \, f_a$, which inside $\bbF_{n_a}$ intersect as $e_a^2 = -n_a$, $f_a^2 = 0$, $e_a \cdot f_a = 1$.
The set of all $(e_a, f_a)$ generate $H_2(Y_6)$.
However, they are not all independent, since the intersections $_{C_a} \overset{\line(2,0){10}}{\;}_{C_{a+1}}$ between $\sigma_a$ and $\sigma_{a+1}$, as indicated in \eqref{eq:surface_config_suN}, impose the gluing condition $C_1 = C_2$ in $Y_6$.\footnote{Note that in \cite{Morrison:2020ool}, the degrees of $\bbF_{n_a}$ are uniformly denoted by $n_a = N-k-2a$ and hence can go negative, with the understanding that in $\bbF_{-n} \cong \bbF_{n}$, the role of $e$ and $h = e+n\,f$ are exchanged \cite{Bhardwaj:2019ngx}.
}

This allows to pick a basis $\{\gamma_i\}$ for $H_2(Y)$, which we set to be
\begin{align}\label{eq:2-cycle_basis_suN}
	\gamma_a = f_a \; (1 \leq a \leq N-1) \, , \quad \gamma_N = e_1 \, .
\end{align}
The other curves $e_a$, $a\geq 2$, are then determined by the gluing conditions $h_a = e_{a-1}$ for $a\leq m$, $h_a = e_{a+1}$ for $a>m$:
\begin{align}\label{eq:gluing_conditions}
	\begin{split}
		e_a & = e_1 - \sum_{b=2}^{a} n_b \, f_b \quad \text{for} \quad 2 \leq a \leq m < N-1\, , \\
		e_{m+1} & = e_m = e_1 - \sum_{b=2}^{m} n_b \, f_b \, , \\
		e_a & = e_1 - \sum_{b=2}^m n_b\,f_b + \sum_{b=m+1}^{a-1} n_b\,f_b \quad \text{for} \quad m+2 \leq a \leq N-1 \, .
	\end{split}
\end{align}
The intersection matrix is
\begin{align}\label{eq:intersection_matrix_suN_5d}
	M_{ai} = \langle \sigma_a, \gamma_i \rangle = 
	\begin{pmatrix}
		-2 & 1 & 0 & \ldots & N-k-4 \\
		1 & -2 & \ddots &  & 2+k - N\\
		0 & \ddots & \ddots & 1 & 0\\
		0 & & 1 & -2 & 0
	\end{pmatrix}
	\equiv 
	\left( \begin{array}{@{}l | r@{}}
		C & \vec{v}
	\end{array} \right) \, ,
\end{align}
where $C_{ab} = \langle \sigma_a, f_b \rangle$ is the (negative) Cartan matrix of $SU(N)$ (see \eqref{eq:suN_cartan_matrix_7d}), and $v_a = \langle \sigma_a, \gamma_N \rangle = \langle \sigma_a , e_1 \rangle$.\footnote{One can infer the intersection pairing $\langle \cdot, \cdot \rangle$ in the Calabi--Yau threefold $Y_6$ from the intersection of 2-cycles $C_1 \cdot C_2$ in $\sigma \cong \bbF_{n}$.
Let $\gamma \subset \sigma$, then $\langle\sigma , \gamma\rangle = -(2e + (n+2)f) \cdot \gamma$.
If there is another 4-cycle $\sigma'$ such that $\sigma$ and $\sigma'$ intersect along $C \subset \sigma$, then $\langle \sigma', \gamma \rangle = C \cdot \gamma$.
}
Note that this intersection matrix does not depend on the different presentation of the surface configuration \eqref{eq:surface_config_suN} compared to \cite{Morrison:2020ool}.
Hence, the subsequent computations of this appendix proceed analogously.

Since $C$ has the known Smith decomposition \eqref{eq:smith_decomp_suN_cartan}, we have
\begin{align}
	M = S \left( \begin{array}{@{}l|r @{}} D & \vec{w} \end{array}  \right) \left(
	\begin{array}{c|c}
		T & 0 \\ \hline
		0 & 1
	\end{array}
	\right) \, , \quad \vec{w} \equiv S^{-1}\vec{v} = (-2, k-N, \ldots, k-N)^t \, ,
\end{align}
where the $(N-1) \times (N-1)$ matrices $(S,D,T)$ are given as in \eqref{eq:smith_decomp_suN_cartan}.
With the first $N-2$ diagonal entries of $D$ being 1, one further can eliminate the first $N-2$ components of $\vec{w}$ with elementary column operations that add multiples of the first $N-2$ columns of $D$ to $\vec{w}$:
\begin{align}
  (D \, | \, \vec{w}) \left( \begin{array}{ccc|cc}
     &  & & 0 & -w_1 \\
     & \mathbbm{1}_{N-2} & & \vdots & \vdots \\
    &  & & 0 & -w_{N-2} \\ \hline
    & 0 & & \multicolumn{2}{c}{\multirow{2}{*}{$\mathbbm{1}_2$}} \\
    & 0 & & 
  \end{array} \right) =
  \begin{pmatrix}
    1 & 0 &  & & 0 \\
    0 & \ddots & & & \vdots \\
    & & 1 & 0 & \vdots \\
    & & & N & k-N
  \end{pmatrix} \equiv ( D' \, | \, \vec{w}') \, .
\end{align}
For the last two entries, $(N, k-N)$, we can use an extended Euclidean algorithm, such that there is an invertible integer $2 \times 2$ matrix $Q^{-1}$ with
\begin{align}\label{eq:Q_matrix_def}
\begin{split}
  & \underbrace{\det(Q)}_{=\pm 1} \left( Q_{22} N - Q_{21} (k-N)), (k-N) Q_{11}-Q_{12} N \right) \equiv (N, k-N) \, Q^{-1} = \det(Q) (\text{gcd}(N,k), 0) \\
  & \Longrightarrow \quad  Q_{11} = \frac{N}{\text{gcd}(N,k)} =: \ell \, , \quad Q_{12} = \frac{k-N}{\text{gcd}(N,k)} =: \tilde\ell \, , \quad Q_{22} = \frac{\det{Q} + \tilde\ell \, Q_{21}}{\ell} \in \bbZ \, .
\end{split}
\end{align}
This can be summarized as
\begin{align}
\begin{split}
	& \left( \begin{array}{@{}l|r @{}} D' & \vec{w}' \end{array}  \right) 
	\left( \begin{array}{ccc|cc}
		 &  & & 0 & 0 \\
		 & \mathbbm{1}_{N-2} & & \vdots & \vdots \\
		&  & & 0 & 0 \\ \hline
		& 0 & & \multicolumn{2}{c}{\multirow{2}{*}{$Q^{-1}$}} \\
		& 0 & & 
	\end{array} \right)
	=
	\begin{pmatrix}
		1 & 0 &  & & 0 \\
		0 & \ddots & & & \vdots \\
		& & 1 & 0 & \vdots \\
		& & & \text{gcd}(N,k) & 0
	\end{pmatrix} 
	\equiv \left( D'' \,  |  \, 0 \right) \, ,
\end{split}
\end{align}
which shows explicitly that $H_4(Y_6, \partial Y_6)/\text{im}(\jmath_4) \cong \bbZ \oplus \bbZ_{\text{gcd}(N,k)}$ \cite{Morrison:2020ool}.
This means that the Smith decomposition of \eqref{eq:intersection_matrix_suN_5d} takes the form
\begin{align}
\begin{split}
	M & = S \, \left( D'' \, | \, 0 \right) 
  \left( \begin{array}{ccc|cc}
     &  & & 0 & 0 \\
     & \mathbbm{1}_{N-2} & & \vdots & \vdots \\
    &  & & 0 & 0 \\ \hline
    & 0 & & \multicolumn{2}{c}{\multirow{2}{*}{$Q^{-1}$}} \\
    & 0 & & 
  \end{array} \right)^{-1}
  \left( \begin{array}{ccc|cc}
     &  & & 0 & -w_1 \\
     & \mathbbm{1}_{N-2} & & \vdots & \vdots \\
    &  & & 0 & -w_{N-2} \\ \hline
    & 0 & & \multicolumn{2}{c}{\multirow{2}{*}{$\mathbbm{1}_2$}} \\
    & 0 & & 
  \end{array} \right)^{-1} \left(\begin{array}{c|c}
		T & 0 \\ \hline
		0 & 1
	\end{array}
	\right) \\
	& = S \, \left( D'' \, | \, 0 \right) \left( \begin{array}{c|c}
		\mathbbm{1}_{N-2} & * \\ \hline
		0 & Q
	\end{array} \right)
	\left( 
		\begin{array}{c|c}
			\mathbbm{1}_{N-2} & *' \\ \hline
			0 & \mathbbm{1}_{2}
		\end{array}
	\right) = S \, \left( D'' \, | \, 0 \right) 
		\underbrace{\left( \begin{array}{c|c}
		\mathbbm{1}_{N-2} & \tilde{*} \\ \hline
		0 & Q
	\end{array} \right)}_{\equiv \tilde{T}} \, ,
\end{split}
\end{align}
where we have used the schematic form \eqref{eq:smith_decomp_suN_cartan} of the $(N-1) \times (N-1)$ matrix $T = \left( \begin{smallmatrix} \mathbbm{1}_{N-2} & \ast' \\ 0 & 1\end{smallmatrix} \right)$.
We are omitting the details on the upper right part of the matrices, because these ultimately specify only the generators of $H_4(Y_6, \partial Y_6)$ that lie in $\text{im}(\jmath_4)$ in terms of the dual basis of $\gamma_i$; these are projected out in the quotient that determines the global and 1-form symmetries.
The decomposition also tells us that the $\text{gcd}(N,k)$-torsional boundary flux may be represented as
\begin{align}
\frac{1}{\text{gcd}(N,k)} \sum_a (S^{-1})_{N-1,a} \, \jmath_4(\sigma_a) = \frac{1}{\text{gcd}(N,k)} \sum_a a \, \jmath_4(\sigma_a) \, .
\end{align}
Moreover, we have the generator
\begin{align}\label{eq:def_top_U1_suN_5d_example}
\begin{split}
	& \epsilon_I = \sum_{i} \tilde{T}_{N,i} \eta_i = Q_{21} \, \eta_{N-1} + Q_{22} \, \eta_{N} \in \text{Hom}(H_2(Y_6), \bbZ) \, : \\
	& \sum_{i=1}^N \mu_i \gamma_i \equiv \sum_{i=1}^{N-2} \mu_i \gamma_i + \mu_{N-1} \, f_{N-1} + \mu_N \, e_1 \mapsto Q_{21} \, \mu_{N-1} + Q_{22} \, \mu_N \, .
\end{split}
\end{align}

Furthermore, the intersections $\gamma_{ab} = \sigma_a \cdot \sigma_b$ with $a \neq b$, as indicated by \eqref{eq:surface_config_suN}, can be expressed in the basis $\gamma_i \in \{f_1, f_2,..., f_{N-1}, e_1\}$ via \eqref{eq:gluing_conditions}:
$\gamma_{ab}$ is either trivial (if $|a-b|>1$), or one of the gluing curves in \eqref{eq:surface_config_suN}, whose homology class takes the form $e_l = e_1 + \sum_{c} (\pm n_c) f_c$.
Omitting the precise expression, the important point becomes that the sum \emph{never} contains $f_{N-1}$ because of \eqref{eq:gluing_conditions}.
For $\gamma_{aa} = \sigma_a \cdot \sigma_a = -(2e_a + n_a f_a) = -2e_1 + \sum_b y_b f_b$, the sum also does not contain $f_{N-1}$ for $a<N-1$.
For $a=N-1$, we have
\begin{align}
	\gamma_{N-1, N-1} = -(2e_{N-1} + (n_{N-1}+2) f_{N-1}) = -2e_1 + \sum_{b=2}^{N-2} y_b' f_b - (N+k) f_{N-1}
\end{align}
for some $y_b'$.
Therefore, we have
\begin{align}
\begin{split}
	K_{Iab} & = \epsilon_I(\sigma_a \cdot \sigma_b) = \begin{cases}
		-2Q_{22} - \delta_{a,N-1} (N+k) Q_{21} \, , & \text{if } \, a = b \, ,\\
		Q_{22} \, , & \text{if } \, b=a+1 \, ,\\
		Q_{22} \, , & \text{if } \, b=a-1 \, ,\\
		0 \, & \text{otherwise} \, .
	\end{cases} \\
	& = Q_{22} \, (-C_{ab}) -  \delta_{a,N-1}\delta_{b,N-1} (N+k) \, Q_{21} \, .
\end{split}
\end{align}
Therefore, in the presence of the $\bbZ_{\text{gcd}(N,k)}$ 1-form symmetry background, the cross-term's coefficients in \eqref{eq:fractional_shift_CS-terms_5d} are
\begin{align}
\begin{split}
	\sum_{a} K_{Iab} \frac{\lambda_a}{\text{gcd}(N,k)} & = \frac{1}{\text{gcd}(N,k)} \sum_a a  \left( Q_{22} \, (-C_{ab})  - \delta_{a,N-1}\delta_{b,N-1} (N+k) Q_{21} \right)\\
	& = 
	\begin{cases}
		0 \, , & \text{if } \, b < N-1 \, , \\
		\frac{Q_{22} \, N - (N-1)(N+k) Q_{21}}{\text{gcd}(N,k)} \, , & \text{if } \, b = N-1 \, ,
	\end{cases}
\end{split}
\end{align}
where we have used \eqref{eq:sum_cartan_a} for the particular weighted sum over Cartan matrix entries of $SU(N)$.
Since both $N$ and $(N+k)$ divide $\text{gcd}(N,k)$, these coefficients are indeed integer.

Finally, we can compute the fractional shift of the instanton density,
\begin{align}
\begin{split}
	& \frac12 \sum_{a,b} \int_{M_5}  K_{Iab} \, A_I \wedge \left( F_a \wedge F_b  + \frac{\lambda_a \lambda_b}{\text{gcd}(N,k)^2} B \wedge B \right) \\
	= & \int_{M_5} \left( \sum_{a,b}  \frac{Q_{22}}{2} (-C_{ab}) A_I \wedge F_a \wedge F_b - \frac{(N+k)Q_{21}}{2} A_I \wedge F_{N-1} \wedge F_{N-1} \right) \\
	+ & \int_{M_5} \frac{1}{2 \, \text{gcd}(N,k)^2} \left( Q_{22} \, N(N-1) - Q_{21} (N+k) (N-1)^2 \right) B \wedge B \, .
\end{split} 
\end{align}
For the last term, we define $\ell = N/\text{gcd}(N,k)$; we can view the $\bbZ_{\text{gcd}(N,k)}\subset \bbZ_{N}$ subgroup of the center of $SU(N)$ being generated by $\ell \mod N$.
Then
\begin{align}
\begin{split}
	& \frac{1}{2\text{gcd}(N,k)^2} \left( Q_{22} \, N(N-1) - Q_{21} (N+k) (N-1)^2 \right) \\
	= &\frac{N-1}{2N} \ell^2 \left( Q_{22}  - Q_{21} (N+k) \frac{N-1}{N} \right) \, .
\end{split}
\end{align}

\end{appendix}


\bibliographystyle{JHEP.bst}
\bibliography{biblio.bib}
\end{document}